\documentclass[preprint,showpacs,aps,floats,nofootinbib]{revtex4}
\usepackage{graphicx}
\usepackage{amsmath}
\usepackage{amssymb}
\begin{document}
\begin{flushright}
\begin{tabular}{l}
 \\
CYCU-HEP-09-12
\end{tabular}
\end{flushright}

\title{
Radiative and Semileptonic $B$ Decays Involving Higher $K$-Resonances in the Final States}
\author{Hisaki Hatanaka}
\author{Kwei-Chou Yang}
\affiliation{Department of Physics, Chung-Yuan Christian University,\\
Chung-Li, Taiwan 320, R.O.C.}
\date{\today}
\begin{abstract}
We study the radiative and semileptonic $B$ decays involving a spin-$J$ resonant $K_J^{(*)}$ with parity $(-1)^J$ for $K_J^*$ and $(-1)^{J+1}$ for $K_J$ in the final state. Using the large energy effective theory (LEET) techniques, we formulate $B \to K_J^{(*)}$ transition form factors in the large recoil region in terms of two independent LEET functions $\zeta_\perp^{K_J^{(*)}}$ and $\zeta_\parallel^{K_J^{(*)}}$, the values of which at zero momentum transfer are estimated in the BSW model. According to the QCD counting rules, $\zeta_{\perp,\parallel}^{K_J^{(*)}}$ exhibit a dipole dependence in $q^2$. We predict the decay rates for $B \to K_J^{(*)} \gamma$, $B \to K_J^{(*)} \ell^+ \ell^-$ and $B \to K_J^{(*)}\nu \bar{\nu}$. The branching fractions for these decays with higher $K$-resonances in the final state are suppressed due to the smaller phase spaces and the smaller values of $\zeta^{K_J^{(*)}}_{\perp,\parallel}$. Furthermore, if the spin of $K_J^{(*)}$ becomes larger, the branching fractions will be further suppressed due to the smaller Clebsch-Gordan coefficients defined by the polarization tensors of the $K_J^{(*)}$. We also calculate the forward backward asymmetry of the $B \to K_J^{(*)} \ell^+ \ell^-$ decay, for which the zero is highly insensitive to the $K$-resonances in the LEET parametrization.
\end{abstract}
\maketitle
%%%%%%%%%%%%%%%%%%%%%%%%%%%%%%%%%%%%%%%
%  Slash character...
%
\newcommand{ \slashchar }[1]{\setbox0=\hbox{$#1$}   % set a box for #1
   \dimen0=\wd0                                     % and get its size
   \setbox1=\hbox{/} \dimen1=\wd1                   % get size of /
   \ifdim\dimen0>\dimen1                            % #1 is bigger
      \rlap{\hbox to \dimen0{\hfil/\hfil}}          % so center / in box
      #1                                            % and print #1
   \else                                            % / is bigger
      \rlap{\hbox to \dimen1{\hfil$#1$\hfil}}       % so center #1
      /                                             % and print /
   \fi}                                             %
%%EXAMPLE:  $\slashchar{E}$ or $\slashchar{E}_{t}$
%
% show equation number including section number...
%\renewcommand{\theequation}{\thesection.\arabic{equation}}
%\makeatletter
%\@addtoreset{equation}{section}
%\makeatother
%
% CAUTION --- bm.sty and amsmath.sty comflict. chose just one of them
%\usepackage{bm}
%\usepackage{amsmath}
%\newcommand{\boldm}[1]{\mbox{\boldmath{${#1}$}}}
%
\newcommand{\Tr}{{\mathop{\mbox{Tr}}\nolimits}} % Trace "Tr"
\newcommand{\tr}{{\mathop{\mbox{tr}}\nolimits}} % trace "tr"
\newcommand{\Det}{{\mathop{\mbox{Det}}\nolimits}} % Determinant "Det"
\newcommand{\diag}{{\mathop{\mbox{diag}}\nolimits}} % diagonal matrix "diag"
\newcommand{\Diag}{{\mathop{\mbox{Diag}}\nolimits}} % diagonal matrix "diag"
\newcommand{\Li}{{\mathop{\mbox{Li}}\nolimits}} % Polylogarithm "Li"
\renewcommand{\Re}{\mathop{\mbox{Re}}} % Real part "Re"
\renewcommand{\Im}{\mathop{\mbox{Im}}} % Imaginary part "Im"
\newcommand{\del}{\partial}
\newcommand{\bvec}[1]{{\boldsymbol{#1}}}
%
% Script symbols
\newcommand{\calA}{{\cal A}}
\newcommand{\calB}{{\cal B}}
\newcommand{\calC}{{\cal C}}
\newcommand{\calD}{{\cal D}}
\newcommand{\calE}{{\cal E}}
\newcommand{\calF}{{\cal F}}
\newcommand{\calG}{{\cal G}}
\newcommand{\calH}{{\cal H}}
\newcommand{\calI}{{\cal I}}
\newcommand{\calJ}{{\cal J}}
\newcommand{\calK}{{\cal K}}
\newcommand{\calL}{{\cal L}}\renewcommand{\L}{{\cal L}}
\newcommand{\calM}{{\cal M}}
\newcommand{\calN}{{\cal N}}
\newcommand{\calO}{{\cal O}}
\newcommand{\calP}{{\cal P}}
\newcommand{\calQ}{{\cal Q}}
\newcommand{\calR}{{\cal R}}
\newcommand{\calS}{{\cal S}}
\newcommand{\calT}{{\cal T}}
\newcommand{\calU}{{\cal U}}
\newcommand{\calV}{{\cal V}}
\newcommand{\calW}{{\cal W}}
\newcommand{\calX}{{\cal X}}
\newcommand{\calY}{{\cal Y}}
\newcommand{\calZ}{{\cal Z}}
\newcommand{\bra}[1]{{\langle {#1}}}
\newcommand{\ket}[1]{{ {#1} \rangle}}
\newcommand{\bbra}[1]{{\langle {#1} |}}
\newcommand{\bket}[1]{{| {#1} \rangle}}
\newcommand{\SM}{{\rm SM}}
\newcommand{\psibar}{\bar{\psi}}
\newcommand{\barpsi}{\bar{\psi}}
\newcommand{\qbar}{\bar{q}}
\newcommand{\ubar}{\bar{u}}
\newcommand{\dbar}{\bar{d}}
\newcommand{\cbar}{\bar{c}}
\newcommand{\sbar}{\bar{s}}
\newcommand{\tbar}{\bar{t}}
\newcommand{\bbar}{\bar{b}}
\newcommand{\nubar}{\bar{\nu}}
\renewcommand{\l}{\ell}
\newcommand{\lbar}{\bar{\l}}
\newcommand{\eV}{{\rm eV}}
\newcommand{\MeV}{{\rm MeV}}
\newcommand{\GeV}{{\rm GeV}}
\newcommand{\TeV}{{\rm TeV}}
\newcommand{\degree}{^\circ}
\newcommand{\rad}{{\rm rad.}}
\newcommand{\Kelvin}{^\circ {\rm K}}
\newcommand{\doK}{^\circ {\rm K}}
\newcommand{\Celsius}{^\circ {\rm C}}
\newcommand{\doC}{^\circ {\rm C}}
\newcommand{\BABAR}{BABAR\ }
\newcommand{\BELLE}{Belle\ }
\newcommand{\eps}{\epsilon}
\newcommand{\veps}{\varepsilon}
\newcommand{\mnrs}{\mu\nu\rho\sigma}
\newcommand{\mnr}{\mu\nu\rho}
\newcommand{\munu}{\mu\nu}
\newcommand{\alphabeta}{\alpha\beta}
\newcommand{\tilden}{\tilde{n}}
\newcommand{\scv}{\slashchar{v}}
\newcommand{\scn}{\slashchar{n}}
\newcommand{\para}{\parallel}

\newcommand{\bars}{\bar{s}}

\newcommand{\zetav}{\zeta^{\knst(v)}}
\newcommand{\zetaa}{\zeta^{\knst(a)}}
\newcommand{\zetat}{\zeta^{\knst(t)}}
\newcommand{\zetatf}{\zeta^{\knst(t_5)}}

\newcommand{\zetavA}{\zeta^{\kn(v)}}
\newcommand{\zetaaA}{\zeta^{\kn(a)}}
\newcommand{\zetatA}{\zeta^{\kn(t)}}
\newcommand{\zetatfA}{\zeta^{\kn(t_5)}}

\newcommand{\hatV}{\hat{V}}
\newcommand{\hatA}{\hat{A}}
\newcommand{\hatT}{\hat{T}}
\newcommand{\hatTf}{{\hat{T}_5}}
\newcommand{\hatS}{\hat{S}}
\newcommand{\hatP}{\hat{P}}
\newcommand{\cdotv}{\cdot v}

\newcommand{\Br}{{\cal B}}
\newcommand{\eff}{{\rm eff}}
\newcommand{\barB}{\overline{B}}
\newcommand{\barBz}{\overline{B}^0}
\newcommand{\mB}{m_{B}}
\renewcommand{\l}{\ell}
\newcommand{\lp}{\l^+}
\newcommand{\lm}{\l^-}
\newcommand{\lpm}{\l^+\l^-}
\newcommand{\mupm}{\mu^+\mu^-}
\newcommand{\kstar}{K^*}
\newcommand{\kstarp}{K^{*+}}
\newcommand{\kstarv}{K^*(892)}
\newcommand{\kstarpv}{K^{*+}(892)}
\newcommand{\knst}{K_{J}^{*}}
\newcommand{\barknst}{\overline{K}_{J}^*}
\newcommand{\barknpst}{\overline{K}_{J}^{(*)}}
\newcommand{\mknst}{m_{\knst}}
\newcommand{\mknpst}{m_{\knpst}}
\newcommand{\tveps}{\tilde{\veps}_{(J)}{}}
\newcommand{\tvepsst}{{\tilde{\veps}_{(J)}^*}{}}
\newcommand{\kn}{K_J}
\newcommand{\barkn}{\overline{K}_J}
\newcommand{\mkn}{m_{\kn}}
\newcommand{\knpst}{K_J^{(*)}}
\newcommand{\knpstp}{K_J^{(*)+}}
\newcommand{\knpstz}{K_J^{(*)0}}
\newcommand{\knpstm}{K_J^{(*)-}}
\newcommand{\barknpstz}{\overline{K}_{J}^{(*)0}}
\newcommand{\knstv}[2]{K_{#1}^*({#2})}
\newcommand{\knstpv}[2]{K_{#1}^{*+}({#1})}
\newcommand{\knv}[2]{K_{#1}({#2})}
\newcommand{\ep}{e}
\newcommand{\alphaem}{\alpha_{EM}}
\newcommand{\hats}{\hat{s}}
\newcommand{\hatu}{\hat{u}}
\newcommand{\hatus}{\hat{u}(s)}
\newcommand{\hatm}{\hat{m}}
\newcommand{\hatmknst}{\hat{m}_{\knst}}
\newcommand{\hatmknpst}{\hat{m}_{\knpst}}
\newcommand{\hatml}{\hat{m}_{\l}}
\newcommand{\hatmb}{\hat{m}_b}
\newcommand{\alphalJ}{\alpha_L^{(J)}{}}
\newcommand{\betatJ}{\beta_T^{(J)}{}}
\newcommand{\alphal}{{\alpha_L}}
\newcommand{\betat}{{\beta_T}}
\newcommand{\Hc}{\mbox{H.c.}}
\newcommand{\fig}{Fig.~}
\newcommand{\tbl}{Table~}
\newcommand{\sect}{Sec.~}
\newcommand{\app}{Appendix}

\newcommand{\err}[2]{{}^{#1}_{#2}{}}
%
%\tableofcontents
%
%%%%%%%%%%%%%%%%%%%%%%%%%%%%%%%%%%%%%%%%%%%%%%%%%%%%%%%%%%%%%%%%%%%%%%%%%%
\section{Introduction}
%%%%%%%%%%%%%%%%%%%%%%%%%%%%%%%%%%%%%%%%%%%%%%%%%%%%%%%%%%%%%%%%%%%%%%%%%%

%%%%%%%%%%%%%%%%%%%%%%%%%%%%%%
\begin{table}[tbp]
\caption{The data for branching ratios of the radiative and semi-leptonic $B$ decays
involving strange mesons.}\label{btosdata}
\begin{ruledtabular}
\begin{tabular}{lclc}
mode & $\Br$ [$10^{-6}$] & mode & $\Br$ [$10^{-6}$]
\\
\hline
$B^+\to K^{*+}(892)\gamma$ & $43.6\pm1.8$ \cite{Coan:1999kh,Nakao:2004th,:2008cy,Aubert:2009we}&
$B^0\to K^{*0}(892)\gamma$ & $43.3\pm1.5$ \cite{Coan:1999kh,Nakao:2004th,:2008cy,Aubert:2009we}
\\
$B^+\to K_2^{*+}(1430)\gamma$ & $14.5\pm4.3$ \cite{Aubert:2003zs}&
$B^0\to K_2^{*0}(1430)\gamma$ & $12.4\pm2.4$ \cite{Aubert:2003zs,Nishida:2002me}
\\
$B^+\to K_3^{*+}(1780)\gamma$ & $<39$ \cite{Nishida:2004fk} &
$B^0\to K_3^{*0}(1780)\gamma$ & $<83$ \cite{Nishida:2004fk}
\\
\hline
$B^+\to K^{*+}(892)e^+ e^-$ & $1.42^{+0.43}_{-0.39}$ \cite{:2008sk,Aubert:2006vb} &
$B^0\to K^{*0}(892)e^+ e^-$ & $1.13^{+0.21}_{-0.18}$ \cite{:2008sk,Aubert:2006vb}
\\
$B^+\to K^{*+}(892)\mu^+ \mu^-$ & $1.12^{+0.32}_{-0.27}$  \cite{:2008sk,Aubert:2006vb} &
$B^0\to K^{*0}(892)\mu^+ \mu^-$ & $1.00^{+0.15}_{-0.13}$  \cite{:2008sk,Aubert:2006vb,Aaltonen:2008xf}
\\
$B^+\to K^{*+}(892)\nu \nubar$ & $ < 80$  \cite{:2008fr,:2007zk}&
$B^0\to K^{*0}(892)\nu \nubar$ & $ < 120$ \cite{:2008fr,:2007zk}
\\
\hline
$B^+ \to K_1^+(1270)\gamma$ & $43\pm12$ \cite{Yang:2004as} &
$B^0 \to K_1^0(1270)\gamma$ & $<58 $ \cite{Yang:2004as}
\\
$B^+ \to K_1^+(1400)\gamma$ & $<15$ \cite{Yang:2004as} &
$B^0 \to K_1^0(1400)\gamma$ & $<15$ \cite{Yang:2004as}
\\
\hline
 $b \to s \gamma$ &  $352  \pm 25$ \cite{Abe:2008sxa,Aubert:2005cua,Chen:2001fja} &
 $b \to s \lpm$   &  $4.50^{+1.03}_{-1.01}$ \cite{Iwasaki:2005sy,Aubert:2004it,Glenn:1997gh}
\\
\end{tabular}
\end{ruledtabular}
\end{table}
%%%%%%%%%%%%%%%%%%%%%%%%%%%%%%%%

The flavor-changing neutral current (FCNC) $b \to s$ processes suppressed in the standard model (SM) could receive sizable new-physics contributions.
Recently \BABAR{} and \BELLE{}  have shown interesting results on the longitudinal fraction, forward-backward asymmetry and isospin asymmetry of the $B \to \kstar\lpm$decays \cite{Ishikawa:2006fh,Aubert:2006vb,:2008ju,Aubert:2008ps,:2008sk,Wei:2009zv}.
Although the data are still consistent with the SM predictions, they favor the flipped-sign $c_7^{\eff}$ models \cite{Eigen:2008nz}.
The minimal flavor violation supersymmetry models with large $\tan\beta$ can be fine-tuned to have the flipped sign $c_7^{\eff}$, where the dominant contributions due to the charged Higgs exchange to $c_9$ and $c_{10}$ are suppressed by $1/\tan^2\beta$ for large $\tan\beta$ \cite{Feldmann:2002iw,Ali:1999mm}. The LHCb is devoted to the $B$ physics studies. Due to the large cross section for $b\bar{b}$ production, the measurement for the rare decays can extend down to $10^{-9}$ branching ratio. It was estimated by the LHCb collaboration that with a data set of 2 fb$^{-2}$ the $B\to K^* \ell^+ \ell^-$ signal events can be improved by an order of magnitude compared with the present results.

Using the large energy effective theory (LEET) techniques \cite{Charles:1998dr},
we have formulated the $B \to K_2^*(1430)$ form factors in the large recoil region \cite{Hatanaka:2009gb},
and further studied the decays $B\to \knstv{2}{1430}\gamma$, $B\to \knstv{2}{1430} \lpm$ and $B \to \knstv{2}{1430}\nu \nubar$. In this paper we will generalize to the studies of  $B\to\knpst\gamma$, $B\to\knpst\l^+\l^-$ and $B\to\knpst\nu\nubar$ decays within the SM,
where $K_J^*$ and $K_J$ are the spin-$J$ resonances with parities $(-1)^J$ and $(-1)^{J+1}$, respectively. We anticipate to see these modes at LHCb, compared with the current data in Table~\ref{btosdata} \cite{Coan:1999kh,Nakao:2004th,:2008cy,Aubert:2003zs,Nishida:2002me,Nishida:2004fk,:2008sk,Aubert:2006vb,:2008fr,:2007zk,Yang:2004as,Abe:2008sxa,Aubert:2005cua,Chen:2001fja,Iwasaki:2005sy,Aubert:2004it,Glenn:1997gh,Amsler:2008zz,Aaltonen:2008xf,Barberio:2008fa,Aubert:2009we}.
In the present study, we will show that the form factors for general $B \to K^{(*)}_J$ transitions can be parametrized in terms of two independent LEET functions, $\zeta_\perp^{K_J^{(*)}}(q^2)$ and $\zeta_\para^{K_J^{(*)}}(q^2)$ together with the Clebsch-Gordan coefficients, $\alpha_L^{(J)}$ and $\beta_T^{(J)}$. The values of $\zeta_\perp^{K_J^{(*)}}(0)$ and $\zeta_\para^{K_J^{(*)}}(0)$ will be estimated by using the Bauer-Stech-Wirbel (BSW) model \cite{BSW}. Moreover, we find that branching fractions with higher resonances, $K_J^{(*)}$, becomes smaller not only due to their smaller phase spaces, but also to the smaller $\zeta^{\knpst}_{\perp,\parallel}$. Meanwhile, the branching fractions involving $K_J^{(*)}$ with higher spin $J$ will be further suppressed due to smaller Clebsch-Gordan coefficients defined by the polarization tensors of the $K_J^{(*)}$.

There have been a few studies of radiative $B$ decays into higher $K$-resonances in the literature \cite{Ali:1992zd,Ebert:2001en,Cheng:2004yj,Hatanaka:2009gb,Hatanaka:2008xj}. A discussion for the general cases was given in Ref.~\cite{Ali:1992zd}, where for various processes the authors parameterize the relevant form factors into four isgur-Wise functions, which are estimated from Isgur-Scora-Grinstein-Wise (ISGW) model \cite{Isgur:1988gb}. However, they obtained ${\cal B}(B\to K_1(1270) \gamma) < {\cal B}(B\to K_1(1400) \gamma)\simeq (2.4 - 5.2)\times 10^{-5}$, in contradiction to the observation (see Table~\ref{btosdata}). One of the motivations for this work is further to re-examine the other radiative decay channels with higer $K$-resonances.

This paper is organized as follows.
In \sect\ref{sec:LEET-FF} we formulate the $B \to \knpst$ form factors using the LEET techniques. In \sect\ref{sec:numerical} we estimate the LEET form factors,  $\zeta_\perp^{K_J^{(*)}}(0)$ and $\zeta_\para^{K_J^{(*)}}(0)$, in the BSW model, and then numerically study the radiative and semileptonic $B$ meson decays into the $\knpst$, including the analyses for the forward-backward asymmetries and longitudinal fraction distributions for $B \to \knpst \mupm$.
We conclude with a summary in \sect\ref{sec:summary}.
The derivation of the $B \to \kn$ form factors is given in Appendix \ref{app:btotn-FF}.

%%%%%%%%%%%%%%%%%%%%%%%%%%%%%%%%%%%%%%%%%%%%%%%%%%%%%%%%%%%%%%%%%%%%%%%%%%
\section{$B \to \knst$ form factors in the large recoil region}\label{sec:LEET-FF}
%%%%%%%%%%%%%%%%%%%%%%%%%%%%%%%%%%%%%%%%%%%%%%%%%%%%%%%%%%%%%%%%%%%%%%%%%%

In this section, using the LEET technique, we formulate $B \to \knst$ form factors in the large recoil region.
The analogous formulation for $B \to \kn$ form factors is given in \app~\ref{app:btotn-FF}.
In this paper $K_J^*$ and $K_J$ stand for the higher spin-$J$ $K$-resonances with parities $(-1)^J$ and $(-1)^{J+1}$, respectively.
For simplicity we study in the rest frame of the $B$ meson (with mass $m_B$)
and assume that the tensor meson $\knst$ (with mass $\mknst$ and energy $E$) moves along the $z$-axis.
In the LEET limit, $E,m_B \gg \mknst, \Lambda_{\rm QCD}$, the momenta of the $B$ and $\knst$ are given by
\begin{eqnarray}
p_B^\mu = (m_B,0,0,0) = m_B \, v^\mu,
\quad
p_{\knst}^\mu = (E,0,0,p_3) \simeq E\, n^\mu,
\end{eqnarray}
respectively. Here $v^\mu = (1,0,0,0)$, $n^\mu=(1,0,0,1)$, and the tensor meson's energy $E$ is given by
\begin{eqnarray}
E = \frac{m_B}{2}\left(1 - \frac{q^2}{m_B^2} + \frac{m_{\knst}^2}{m_B^2}\right),
\end{eqnarray}
with $q=p_B - p_{\knst}$.

The polarization tensors $\veps(\lambda)^{\mu_1 \mu_2 \cdots \mu_J}$ of the massive spin-$J$ meson with helicity $\lambda$ that can be constructed in terms of the polarization vectors of a massive vector state with the mass $m_{\knst}$
\begin{eqnarray}
\veps(0)^{*\mu} = (p_3,0,0,E)/\mknst,
\quad
\veps(\pm1)^{*\mu} = (0,\mp1,+i,0)/\sqrt{2},
\end{eqnarray}
are given by
\begin{eqnarray}
\veps(\pm2)^{\mu\nu} &\equiv& \veps(\pm1)^\mu \veps(\pm1)^\nu,
\\
\veps(\pm1)^{\mu\nu} &\equiv& \sqrt{\frac{1}{2}}
\left[ \veps(\pm1)^\mu \veps(0)^\nu + \veps(0)^\mu \veps(\pm1)^\nu \right],
\\
\veps(0)^{\mu\nu}&\equiv& \sqrt{\frac{1}{6}}
 \left[ \veps(+1)^\mu \veps(-1)^\nu  + \veps(-1)^\mu \veps(+1)^\nu \right]
  + \sqrt{\frac{2}{3}} \veps(0)^\mu \veps(0)^\nu,
\end{eqnarray}
for $J=2$ and
\begin{eqnarray}
\veps(\pm3)^{\mnr} &=& \veps(\pm1)^{\mu}\veps(\pm1)^{\nu}\veps(\pm1)^{\rho},
\\
\veps(\pm2)^{\mnr} &=& \sqrt{\frac{1}{3}} [
 \veps(0)^{\mu}\veps(\pm1)^{\nu}\veps(\pm1)^{\rho} +
 \veps(\pm1)^{\mu}\veps(0)^{\nu}\veps(\pm1)^{\rho} +
 \veps(\pm1)^{\mu}\veps(\pm1)^{\nu}\veps(0)^{\rho}
],
\\
\veps(\pm1)^{\mnr} &=& \sqrt{\frac{1}{15}} [
 \veps(\mp1)^{\mu}\veps(\pm1)^{\nu}\veps(\pm1)^{\rho} +
 \veps(\pm1)^{\mu}\veps(\mp1)^{\nu}\veps(\pm1)^{\rho} +
 \veps(\pm1)^{\mu}\veps(\pm1)^{\nu}\veps(\mp1)^{\rho}
]
\nonumber\\&&+2
\sqrt{\frac{1}{15}} [
 \veps(\pm1)^{\mu}\veps(0)^{\nu}\veps(0)^{\rho} +
 \veps(0)^{\mu}\veps(0)^{\nu}\veps(\pm1)^{\rho} +
 \veps(0)^{\mu}\veps(\pm1)^{\nu}\veps(0)^{\rho}
],
\\
\veps(0)^{\mnr} &=&
\sqrt{\frac{1}{10}} [
 \veps(0)^{\mu}\veps(+1)^{\nu}\veps(-1)^{\rho} +
 \veps(+1)^{\mu}\veps(0)^{\nu}\veps(-1)^{\rho} +
 \veps(+1)^{\mu}\veps(-1)^{\nu}\veps(0)^{\rho} +
\nonumber\\&&\phantom{MMi}
 \veps(0)^{\mu}\veps(-1)^{\nu}\veps(+1)^{\rho} +
 \veps(-1)^{\mu}\veps(0)^{\nu}\veps(+1)^{\rho} +
 \veps(-1)^{\mu}\veps(+1)^{\nu}\veps(0)^{\rho} ] +
\nonumber\\&&
\sqrt{\frac{2}{5}}
 \veps(0)^{\mu}\veps(0)^{\nu}\veps(0)^{\rho},
\end{eqnarray}
for $J=3$, and so on.
$\veps(\lambda)^{\mu_1 \mu_2 \cdots \mu_J}$ is symmetric under interchange of any two of $\mu_j$ and $\mu_k$ ($1\le j,k \le J$), and satisfies
divergence-free conditions
$p_{\knst,\mu} \veps(\lambda)^{\mu\mu_1 \cdots \mu_{J-1}} = 0$,
traceless conditions
$g_{\mu_1 \mu_2} \veps(\lambda)^{\mu_1 \mu_2 \nu_1 \cdots \nu_{J-2}} = 0$,
and orthonormal conditions $\veps(h_1)^{\mu_1\mu_2 \cdots \mu_{J}} \veps(h_2)^*_{\mu_1\mu_2 \cdots \mu_J} = \delta_{h_1 h_2}$.

In the following, we calculate the $\barB \to \barknst$ transition form factors:
\begin{eqnarray}
\bra{\barknst}|V^\mu|\ket{\barB}, \quad
\bra{\barknst}|A^\mu|\ket{\barB}, \quad
\bra{\barknst}|T^{\munu}|\ket{\barB}, \quad
\bra{\barknst}|T_A^{\munu}|\ket{\barB}, \quad
\end{eqnarray}
where $V^\mu = \bars \gamma^\mu b$,
$A^\mu = \bars \gamma^\mu \gamma_5 b$,
$T^{\munu} = \bars \sigma^{\munu} b$ and
$T_A^{\munu} = \bars \sigma^{\munu}\gamma_5 b$.
In the LEET limit
one can easily write down the relevant form factors in terms of
the following projectors
\begin{eqnarray}
(\betatJ)^{-1} \left(\frac{\mknst}{E}\right)^{J-1}
[e(\lambda)^{*\mu} - (e(\lambda)^* \cdot v) n^\mu] &=&
\begin{cases}
0 & \mbox{for } \lambda = \pm2, \\
\veps(\pm1)^{*\mu} & \mbox{for } \lambda = \pm1, \\
0 & \mbox{for } \lambda = 0,
\end{cases} \label{1st-vec}
\\
(\betatJ)^{-1} \left(\frac{\mknst}{E}\right)^{J-1} \eps^{\mnrs}
e(\lambda)^*_\nu n_\rho v_\sigma
&=&\begin{cases}
0 & \mbox{for } \lambda = \pm2, \\
\eps^{\mnrs} \veps(\pm1)^*_\nu n_\rho v_\sigma  & \mbox{for } \lambda = \pm1, \\
0 & \mbox{for } \lambda = 0,
\end{cases} \label{2nd-vec}
\\
(\alphalJ)^{-1} \left(\frac{\mknst}{E}\right)^J  (e(\lambda)^* \cdot v) n^\mu
&=&\begin{cases}
0 & \mbox{for } \lambda = \pm2, \\
0 & \mbox{for } \lambda = \pm1, \\
n^\mu  & \mbox{for } \lambda = 0,
\end{cases} \label{3rd-vec}
\\
(\alphalJ)^{-1}\left(\frac{\mknst}{E}\right)^J (e(\lambda)^* \cdot v) v^\mu
&=&\begin{cases}
0 & \mbox{for } \lambda = \pm2, \\
0  & \mbox{for } \lambda = \pm1, \\
v^\mu  & \mbox{for } \lambda = 0,
\end{cases} \label{4th-vec}
\end{eqnarray}
together with $\eps^{\mu\nu\alpha\beta}$, $v^\mu$ and $n^\mu$,
to project the relevant polarization states of the higher $K$-resonances, where Eqs. \eqref{1st-vec}, \eqref{3rd-vec} and \eqref{4th-vec} are the vectors, but Eq.~\eqref{2nd-vec} the axial-vector.
Here $\veps^{0123}=-1$ and we have defined
\begin{eqnarray}
e(\lambda)^{*\mu} &\equiv& \veps(\lambda)^{*\mu \nu_1 \nu_2 \cdots \nu_{J-1}}
v_{\nu_1} v_{\nu_2} \cdots v_{\nu_{J-1}}
=
\begin{cases}
\displaystyle
 \alphalJ \veps(0)^\mu  \left(\frac{p_3}{\mknst}\right)^{J-1}
 & \mbox{ for } \lambda=0,
\\
\displaystyle
 \betatJ \veps(\pm1)^\mu \left(\frac{p_3}{\mknst}\right)^{J-1}
 & \mbox{ for } \lambda=\pm1,
\end{cases}
\end{eqnarray}
where $\alphalJ$ and $\betatJ$ are the Clebsch-Gordan coefficients of the specific terms of the
polarization tensors:
\begin{eqnarray}
\veps(0)^{\mu\nu_1 \cdots \nu_n} &=& \alphalJ \veps(0)^\mu \veps(0)^{\nu_1} \cdots \veps(0)^{\nu_{J-1}} + \mbox{others},
\\
\veps(\pm1)^{\mu\nu_1 \cdots \nu_n} &=& \betatJ \veps(\pm1)^\mu \veps(0)^{\nu_1} \cdots \veps(0)^{\nu_{J-1}} + \mbox{others},
\end{eqnarray}
and are given by
\newcommand{\cg}[3]{{\calJ}^{(#1)}_{(#2)(#3)}}
\begin{eqnarray}
\alphalJ &=& \cg{J,0}{1,0}{J-1,0} \cg{J-1,0}{1,0}{J-2,0} \cdots \cg{2,0}{1,0}{1,0}
,\\
\betatJ &=& \cg{J,1}{1,1}{J-1,0} \cg{J-1,0}{1,0}{J-2,0} \cg{J-2,0}{1,0}{J-3,0} \cdots \cg{2,0}{1,0}{1,0},
\end{eqnarray}
with $\cg{J,M}{j_1,m_1}{j_2,m_2}$ being the short-hand notations of the following Clebsch-Gordan coefficients
\begin{eqnarray}
\cg{J,M}{j_1,m_1}{j_2,m_2} \equiv \bra{(j_1 m_1),(j_2 m_2)}|\ket{J M}.
\label{CGdef}
\end{eqnarray}
The values of $\alphalJ$ and $\betatJ$ for $J=1,2,\cdots,5$ are collected in \tbl \ref{alphabeta}.

\begin{table}[tbp]
\caption{The Clebsch-Gordan coefficients, $\alphalJ$ and $\betatJ$, with $J=1,2,\cdots,5$.}\label{alphabeta}
\begin{ruledtabular}
\begin{tabular}{lccccc}
$J$ & 1 & 2 & 3 & 4 & 5 %& 6 & 7
\\
\hline
$\alphalJ \phantom{\frac{\dfrac{1}{1}}{1}}$
& $1$
&$\displaystyle\sqrt{\frac{2}{3}}$ %2
&$\displaystyle\sqrt{\frac{2}{5}}$ %3
&$2\sqrt{\dfrac{2}{35}}$ %4
&$\dfrac{2}{3}\sqrt{\dfrac{2}{7}}$ %5
%$\dfrac{4}{\sqrt{231}}$ & %6
%$\dfrac{4}{\sqrt{429}}$  %7
\\
$\betatJ \phantom{\frac{\dfrac{1}{1}}{1}}$
& $1$
&$\sqrt{\dfrac{1}{2}}$  %2
&$\dfrac{2}{\sqrt{15}}$  %3
&$\dfrac{1}{\sqrt{7}}$   %4
&$2\sqrt{\dfrac{2}{105}}$   %5
%$\dfrac{2}{3\sqrt{11}}$ &  %6
%$\dfrac{8}{\sqrt{3003}}$    %7
\end{tabular}
\end{ruledtabular}
\end{table}

Matching the parities of the matrix elements and using the mentioned Lorentz structures, we can then easily parameterize the form factors to be
\begin{eqnarray}
\bra{\barknst}|V^\mu|\ket{\barB}
&=& -i 2 E \left( \frac{\mknst}{E} \right)^{J-1} \zetav_\perp \eps^{\mnrs}
 v_\nu n_\rho e^*_\sigma,
\label{LEETFF-vector}
\\
%%%%%%%%%%%%%%%%%%%%%%%%%%%%
\bra{\barknst}|A^\mu|\ket{\barB}
&=& 2E \left(\frac{\mknst}{E}\right)^{J-1} \zetaa_\perp
\left[ e^{*\mu} - (e^* \cdot v) n^\mu \right]
\nonumber\\&&
+ 2E \left(\frac{\mknst}{E}\right)^{J} (e^* \cdot v)
 \left[\zetaa_{\para} n^\mu + \zetaa_{\para,1} v^\mu \right],
\\
%%%%%%%%%%%%%%%%%%%%%%%%%%%%
\bra{\barknst}|T^{\mu\nu}|\ket{\barB}
&=& 2E \left(\frac{\mknst}{E}\right)^{J}  \zetat_\para (e^* \cdot v)
 \eps^{\mnrs} v_\rho n_\sigma
\nonumber\\&&
+2E \left(\frac{\mknst}{E}\right)^{J-1}\zetat_\perp \eps^{\mnrs} n_\rho [e^*_\sigma
 - (e^* \cdot v) n_\sigma]
\nonumber\\&&
+2E \left(\frac{\mknst}{E}\right)^{J-1}\zetat_{\perp,1} \eps^{\mnrs} v_\rho [e^*_\sigma
 - (e^* \cdot v) n_\sigma],
\label{LEETFF-tensor}
\end{eqnarray}
%%%%%%%%%%%%%%%%%%%%%%%%%%%%
\begin{eqnarray}
\bra{\barknst}|T_A^{\mu\nu}|\ket{\barB}
&=& - i2E \left(\frac{\mknst}{E}\right)^{J-1} \zetatf_{\perp,1}
 \left\{
 \left[ e^{*\mu} - (e^* \cdot v)n^\mu\right] v^\nu - (\mu \leftrightarrow \nu)
 \right\}
\nonumber\\&&
-i2E \left(\frac{\mknst}{E}\right)^{J-1} \zetatf_{\perp} \left\{
 [e^{*\mu} - (e^* \cdot v)n^\mu] n^\nu - (\mu \leftrightarrow \nu)
\right\}
\nonumber\\&&
-i2E \left(\frac{\mknst}{E}\right)^{J} \zetatf_{\para}
(e^* \cdot v) (n^\mu v^\nu - n^\nu v^\mu).
\label{LEETFF-axialtensor}
\end{eqnarray}
Note that the parity of the $\knst$ is $(-1)^J$.
$\bra{\barknst}|T^{\mu\nu}|\ket{\barB}$ is related to $\bra{\barknst}|T_A^{\mu\nu}|\ket{\barB}$ by the relation: $\sigma^{\mu\nu}\gamma_5 \eps_{\mnrs}= 2i \sigma^{\rho\sigma}$.
Note also that  only the $\knst$ with polarization helicities $\pm1$ and $0$ contribute to the $\barB \to \barknst$ transition in the LEET limit, where
$\zeta_\perp$'s are relevant to $\knst$ with helicity = $\pm1$, and $\zeta_\para$'s to $\knst$ with helicity $=0$.

We can further reduce the number for the $\barB \to \barknst$ form factors which are independent,
using the effective current operator $\sbar_n \Gamma b_v$ (with $\Gamma = 1,\gamma_5,\gamma^\mu,\gamma^\mu\gamma_5,\sigma^{\mu\nu},\sigma^{\mu\nu}\gamma_5$) in the LEET limit, instead of the the original one $\sbar \Gamma b$ \cite{Charles:1998dr}.
Here $b_v$ and $s_n$ satisfy $\slashchar{v} b_v = b_v$, $\slashchar{n} s_n = 0$ and $(\slashchar{n}\slashchar{v}/2)s_n = s_n$.
Employing the Dirac identities
\begin{eqnarray}
\frac{\slashchar{v}\slashchar{n}}{2} \gamma^\mu
 &=& \frac{\slashchar{v}\slashchar{n}}{2} \left(n^\mu \slashchar{v} - i \eps^{\mnrs}v_\nu n_\rho \gamma_\sigma \gamma_5 \right),
\\
\frac{\slashchar{v}\slashchar{n}}{2} \sigma^{\mu\nu}
 &=& \frac{\slashchar{v}\slashchar{n}}{2} \left[i(n^\mu v^\nu - n^\nu v^\mu) -i(n^\mu \gamma^\nu - n^\nu \gamma^\mu) \slashchar{v} - \eps^{\mnrs}v_\nu n_\rho \gamma_\sigma \gamma_5 \right],
\end{eqnarray}
one can easily obtain the following relations:
\begin{eqnarray}
\sbar_n b_v &=& v_\mu \sbar_n \gamma^\mu b_v, \label{relation-scalar}
\\
\sbar_n \gamma^\mu b_v &=& n^\mu \sbar_n b_v - i \eps^{\mnrs} v_\nu n_\rho \sbar_n \gamma_\sigma \gamma_5 b_v,
\\
\sbar_n \gamma^\mu\gamma_5 b_v &=&
  - n^\mu \sbar_n \gamma_5 b_v
  - i \eps^{\mnrs}  v_\nu n_\rho \sbar_n \gamma_\sigma b_v,
\\
\sbar_n \sigma^{\mu\nu} b_v &=& i \left[
 n^\mu v^\nu \sbar_n b_v
 - n^\mu \sbar_n \gamma^\nu b_v
 - (\mu \leftrightarrow \nu) \right]
 - \eps^{\mnrs} v_\rho n_\sigma \sbar_n \gamma_5 b_v,
\\
\sbar_n \sigma^{\mu\nu}\gamma_5 b_v &=& i \left[
  n^\mu v^\nu \sbar_n \gamma_5 b_v
  + n^\mu \sbar_n \gamma^\nu \gamma_5 b_v
  - (\mu \leftrightarrow \nu) \right]
  -\eps^{\mnrs} v_\rho n_\sigma \sbar_n b_v. \label{relation-axialtensor}
\end{eqnarray}
We can then obtain
\begin{eqnarray}
\zetav_\perp = \zetaa_\perp = \zetat_\perp = \zetatf_\perp &\equiv& \zeta_\perp^{\knst}(q^2),
\\
\zetaa_\para = \zetat_\para = \zetatf_\para &\equiv& \zeta_\para^{\knst}(q^2),
\\
\zetaa_{\para,1} = \zetatf_{\perp,1} = \zetat_{\perp,1} &=& 0.
\end{eqnarray}
Thus we find that there are only two independent form factors, $\zeta^{\knst}_\perp(q^2)$ and $\zeta^{\knst}_{\para}(q^2)$,
for the $\barB \to \barknst$ transition in the large recoil region.
In the full theory, the $\barB \to \barknst$ form factors are defined as
\begin{eqnarray}
\bra{\barknst(p_{\knst},\lambda)}|\sbar \gamma^\mu b|\ket{\barB(p_B)}
&=& - i \frac{2}{m_B + \mknst} \tilde V^{\knst}(q^2)
\eps^{\mnrs} p_{B\nu} p_{\knst\rho} \ep(\lambda)^*_\sigma,
\label{ff-vector}
\\
%%%%%%%%%%%%
\bra{\barknst(p_{\knst},\lambda)}|\sbar \gamma^\mu \gamma_5 b|\ket{\barB(p_B)}
&=& 2 \mknst \tilde A_0^{\knst}(q^2) \frac{\ep(\lambda)^* \cdot p_B}{q^2} q^\mu
\nonumber\\&&
+ \left(m_B + \mknst\right) \tilde A_1^{\knst}(q^2)
\left[ \ep(\lambda)^{*\mu} - \frac{\ep(\lambda)^* \cdot p_B}{q^2} q^\mu \right]
\nonumber\\&&
- \tilde A_2^{\knst}(q^2) \frac{\ep(\lambda)^* \cdot p_B}{m_B + \mknst}
\left[ p_B^\mu + p_{\knst}^\mu - \frac{m_B^2 - \mknst^2}{q^2} q^\mu
\right],\nonumber\\
\\
%%%%%%%%%%%%
\bra{\barknst(p_{\knst},\lambda)}|\sbar \sigma^{\mu\nu}q_\nu b|\ket{\barB(p_B)}
&=&
 -2 \tilde T_1^{\knst} (q^2) \eps^{\mnrs} p_{B\nu} p_{\knst\rho} \ep(\lambda)^*_\sigma,
\\
%%%%%%%%%%%%
\bra{\barknst(p_{\knst},\lambda)}|\sbar \sigma^{\mu\nu}\gamma_5 q_\nu b|\ket{\barB(p_B)}
&=& -i \tilde T_2^{\knst} (q^2) \biggl[ \left(m_B^2 - \mknst^2\right) \ep(\lambda)^{*\mu}
 \nonumber\\&& \phantom{MMMMM}
 - \left(\ep(\lambda)^* \cdot p_B\right)\left(p_B^\mu + p_{\knst}^\mu\right)
\biggr]
\nonumber\\&&
- i \tilde T_3^{\knst}(q^2) \left(\ep(\lambda)^* \cdot p_B\right)
\left[ q^\mu - \frac{q^2}{m_B^2 - \mknst^2}\left(p_B^\mu + p_{\knst}^\mu\right)
\right],
\nonumber\\ \label{ff-tensor}
\end{eqnarray}
where
\begin{eqnarray}
e(\lambda)^{*\mu} &\equiv& \veps(p_{\knst},\lambda)^{*\mu\nu_1\nu_2 \cdots \nu_{J-1}}
 p_{B,\nu_1}  p_{B,\nu_2}\cdots p_{B,\nu_{J-1}} / m_B^{J-1},
\quad \lambda=0,\pm1.
\end{eqnarray}

Comparing Eqs.~\eqref{ff-vector}-\eqref{ff-tensor} with Eqs.~\eqref{LEETFF-vector}-\eqref{LEETFF-axialtensor}, we obtain
\newcommand{\absp}{|\vec{p}_{\knst}|}
\begin{eqnarray}
\tilde A_0^{\knst}(q^2) \left(\frac{\absp}{\mknst}\right)^{J-1} &\equiv&
 A_0^{\knst}(q^2)
\simeq \left(1-\frac{\mknst^2}{m_B E}\right) \zeta_{\para}^{\knst}(q^2)
+ \frac{\mknst}{m_B} \zeta_{\perp}^{\knst}(q^2),
\label{LEETrelationA0}
\\
\tilde A_1^{\knst}(q^2) \left(\frac{\absp}{\mknst}\right)^{J-1} &\equiv&
 A_1^{\knst}(q^2)
\simeq \frac{2E}{m_B + \mknst} \zeta_{\perp}^{\knst}(q^2),
\\
\tilde A_2^{\knst}(q^2) \left(\frac{\absp}{\mknst}\right)^{J-1} &\equiv&
 A_2^{\knst}(q^2)
\simeq \left(1+\frac{\mknst}{m_B}\right) \left[
\zeta_{\perp}^{\knst}(q^2) -  \frac{\mknst}{E} \zeta_\para^{\knst}(q^2)
\right],
\\
\tilde V^{\knst}(q^2) \left(\frac{\absp}{\mknst}\right)^{J-1} &\equiv&
 V^{\knst}(q^2)
\simeq \left(1+\frac{\mknst}{m_B}\right) \zeta_{\perp}^{\knst}(q^2),
\end{eqnarray}

\begin{eqnarray}
\tilde T_1^{\knst}(q^2) \left(\frac{\absp}{\mknst}\right)^{J-1} &\equiv&
 T_1^{\knst}(q^2)
\simeq \zeta_{\perp}^{\knst}(q^2),
\\
\tilde T_2^{\knst}(q^2) \left(\frac{\absp}{\mknst}\right)^{J-1} &\equiv&
 T_2^{\knst}(q^2)
\simeq \left(1-\frac{q^2}{m_B^2 - \mknst^2}\right)\zeta_{\perp}^{\knst}(q^2),
\\
\tilde T_3^{\knst}(q^2) \left(\frac{\absp}{\mknst}\right)^{J-1} &\equiv&
 T_3^{\knst}(q^2)
\simeq  \zeta_{\perp}^{\knst}(q^2) - \left(1-\frac{\mknst^2}{m_B^2}\right)
\frac{\mknst}{E} \zeta_\para^{\knst}(q^2),
\label{LEETrelationT3}
\end{eqnarray}
where we have used $e^{*\mu} \approx (p_{\knst} / \mknst)^{J-1} \tvepsst^{\mu}$
with $\tveps(0)^\mu = \alphalJ \veps(0)^\mu$,
 $\tveps(\pm1) = \betatJ \veps(\pm1)^\mu$ and $|\vec{p}_{\knst}|/E \simeq 1$.

With the replacement $\veps^\mu \to \tveps^\mu$,
we can easily generalize the studies for $B \to \kstar \gamma$,
$B \to \kstar \lpm$ and $B \to \kstar \nu \nubar$
to the corresponding decays involving resonant strange tensor mesons.

%%%%%%%%%%%%%%%%%%%%%%%%%%%%%%%%%%%%%%%%%%%%%%%%%%%%%%%%%%%%%%%%%%%%%%%%%%%%%%%%%%%%%%%
\section{Numerical Analysis}\label{sec:numerical}
%%%%%%%%%%%%%%%%%%%%%%%%%%%%%%%%%%%%%%%%%%%%%%%%%%%%%%%%%%%%%%%%%%%%%%%%%%%%%%%%%%%%%%%

The properties of $\knpst$ mesons are summarized in \tbl\ref{knstmass}.
%%%%%%%%%%%%%%%%%%%%%%%%%%%%%%%%%%%
\newcommand{\disp}{\text{${}^{(\dag)}$}}
\begin{table}[tbp]
\caption{%
Properties of resonant $\knpst$ mesons (with $J=1,\cdots,5$) \cite{Amsler:2008zz}, and $B\to \knpst$ LEET form factors calculated in the BSW model \cite{BSW}.
$K_1(1270)$ and $K_1(1400)$ are not considered in this paper (see Refs.~\cite{Hatanaka:2008gu,Hatanaka:2008xj}).
States denoted by ``\disp{}'' or ``?'' are not yet well confirmed.
In the present paper we do not take into account $1\,{}^3G_3$ and $1\,{}^3H_4$ states.
}.\label{knstmass}
\begin{ruledtabular}
\begin{tabular}{lccccc}
$\knpst$ & $J^{PC}$ & $n\,{}^{2S+1}L_J$ & $m_{\knpst}$ [\MeV] &$\zeta_\perp(0)$ & $\zeta_\parallel(0)$\\
\hline
$\knstv{}{1410}$        &$1^{--}$& $2\,{}^{3}S_1$? & $1,414\pm15$ & $0.28\pm 0.04$   & $0.22 \pm 0.03$
\\
$\knstv{}{1680}$        &$1^{--}$& $1\,{}^{3}D_1$& $1,717\pm32$   & $0.24\pm 0.05$   & $0.18\pm 0.03$
\\
$\knstv{2}{1430}$       &$2^{++}$& $1\,{}^{3}P_2$& $1,425.6\pm1.5$ ($K_2^{*\pm}$) & $0.28\pm 0.04$  & $0.22\pm 0.03$
\\
                        &        &               & $1,432.4\pm1.3$ ($K_2^{*0}$)   & $0.28\pm 0.04$  & $0.22\pm 0.03$
\\
$\knstv{2}{1980}$ \disp &$2^{+?}$& $1\,{}^{3}F_2$ or $2\,{}^3P_2$? & $1,973 \pm 26$ &$0.20\pm 0.05$ & $0.14\pm 0.03$
\\
$\knstv{3}{1780}$       &$3^{--}$& $1\,{}^{3}D_3$ & $1,776\pm7$     & $0.23\pm 0.05$     & $0.16\pm 0.03$
\\
$\knstv{4}{2045}$       &$4^{++}$& $1\,{}^{3}F_4$ & $2,045\pm9$     & $0.19\pm 0.05$     & $0.13\pm 0.03$
\\
$\knstv{5}{2380}$ \disp & $5^{-?}$& $1\,{}^{3}G_5$? & $2,382\pm24$  & $0.15\pm 0.05$     & $0.10\pm 0.03$
\\
$\knv{1}{1650}$ \disp & $1^{+?}$& $2\,{}^{1}P_1$ or $2\,{}^3P_1$? &  $1,650\pm50 $ & $0.24\pm 0.05$    & $0.18\pm 0.03$
\\
%$\knv{2}{1580}$ \disp & $2^{-?}$& ? & $\sim 1,580$
%\\
$\knv{2}{1770}$       & $2^{-+}$& $1\,{}^1D_2$& $1,773 \pm8$     &  $0.23\pm 0.05$     & $0.17\pm 0.03$
\\
$\knv{2}{1820}$       & $2^{--}$ & $1\,{}^3D_2$? & $1,816\pm13$  &  $0.22\pm 0.05$     & $0.16\pm 0.03$
\\
$\knv{2}{2250}$ \disp &$2^{-?}$& $2\,{}^1D_2$& $2,247\pm17$      &  $0.16\pm 0.05$     & $0.11\pm 0.03$
\\
$\knv{3}{2320}$  \disp & $3^{+?}$& $1\,{}^1F_3$ or $1\,{}3F_3$? & $2,324\pm24$  & $0.15\pm 0.05$    & $0.10\pm 0.03$
\\
$\knv{4}{2500}$  \disp & $4^{-?}$& $1\,{}^1G_4$ or $1^3G_4$? & $2,490\pm20$     & $0.13\pm 0.04$   & $0.09\pm 0.03$
\\
$\knv{5}{2600?}$ \disp & $5^{+?}$& $1\,{}^1H_5$ or $1\,{}^3H_5$? & $\sim2,600?$ & $0.12\pm 0.04$   & $0.08\pm 0.02$
\end{tabular}
\end{ruledtabular}
\end{table}
%%%%%%%%%%%%%%%%%%%%%%%%%%%%%%%%%%
In the following numerical study, we use the values of the parameters
 listed in \tbl\ref{tbl:inputs}.
\begin{table}[tbp]
\caption{Input parameters}\label{tbl:inputs}
\begin{ruledtabular}
\begin{tabular}{ll}
$B$ lifetime (picosecond)&
$\tau_{B^+} = 1.638$,
\quad
$\tau_{B^0} = 1.530$
\\
$b$ quark mass & $m_{b,{\rm pole}} = 4.79^{+0.19}_{-0.08}\, \GeV$
\\
CKM parameter \cite{ckmfitter}
& $|V_{ts}^* V_{tb}^{}| = 0.040\pm0.001$
%\quad $|V_{ub}| = (3.44^{+0.22}_{-0.17}) \times 10^{-3}$.
\end{tabular}
\end{ruledtabular}
\end{table}
%%%%%%%%%%%%%%%%%%%%%%%%%%%%%%%%%%%

%%%%%%%%%%%%%%%%%%%%%%%%%%%%%%%%%%%%%%%%%%%%%%%%%%%%%%%%%%%%%%%%%%%%%%%
\subsection{The determination of form factors and $B\to\knpst\gamma$ Decays}
%%%%%%%%%%%%%%%%%%%%%%%%%%%%%%%%%%%%%%%%%%%%%%%%%%%%%%%%%%%%%%%%%%%%%%%

The $B\to \knpst \gamma$ decay widths are given by
\begin{eqnarray}
\Gamma(B\to\knpst\gamma)
&=&
\frac{G_F^2 \alphaem\left|V_{ts}^* V_{tb}^{}\right|^2}{32\pi^4}
m_{b,{\rm pole}}^2 \mB^3 \left(1- \frac{\mknpst^2}{\mB^2}\right)^3
\nonumber\\&& \phantom{MM}\times
\left|c_7^{(0)\eff} + A^{(1)}(\mu) \right|^2
\left|T_1^{\knpst}(0)\right|^2 \left(\betatJ\right)^2.
\end{eqnarray}
As for the case with $J=2$,
taking into account the data of $\Br(\barB^0 \to \overline{K}_2^{0}\gamma)$ and using
 $c_7^{(0)\eff} = -0.315$, $A^{(1)} = A_{c_7}^{(1)} + A_{\rm ver}^{(1)} = -0.038 - 0.016i$ \cite{Ali:2001ez}, we have obtained \cite{Hatanaka:2009gb}
\begin{eqnarray}
T_1^{K_2^*(1430)}(0) \simeq \zeta_\perp^{K_2^*(1430)}(0)
&=& 0.28 \pm 0.03^{+0.00}_{-0.01},
\label{zetaTvalue}
\end{eqnarray}
where the first and second errors are due to uncertainties of the data and the pole mass of the $b$ quark, respectively.
In the present paper we use the BSW model \cite{BSW} to estimate the LEET form factors at zero momentum transfer, which are be written by
\begin{eqnarray}
\zeta_\perp^{K_J^{(*)}}(0) &=& \frac{m_b-m_s}{2E}J \,,\nonumber\\
\zeta_\parallel^{K_J^{(*)}}(0) &=& \left( A_0^{K_J^{(*)}}(0) - \frac{m_{K_J^{(*)}}}{m_B}\zeta_\perp^{K_J^{(*)}}(0) \right) \left( 1-\frac{m_{K_J^{(*)}}^2}{m_b E} \right)^{-1}\,,
\end{eqnarray}
where, after integrating out the degrees of freedom of the spins,
\begin{eqnarray}
J &=& \sqrt{2}\int d^2p_{\rm T} \int_0^1 \frac{dx}{x}
 \mathbf{\Phi}_{K_J^{(*)}}(p_{\rm T},x)
 \mathbf{\Phi}_{m_B}(p_{\rm T},x) \,, \nonumber\\
A_0^{K_J^{(*)}}(0)&=& \int d^2p_{\rm T} \int_0^1 dx
 \mathbf{\Phi}_{K_J^{(*)}}(p_{\rm T},x)
 \mathbf{\Phi}_{m_B}(p_{\rm T},x) \,.
\end{eqnarray}
Here, for a meson with mass {\it m} its wave function can be parameterized as
\begin{eqnarray}
\mathbf{\Phi}_{m}(p_{\rm T},x) = N_m \sqrt{x(1-x)}
 e^{-p_{\rm T}^2 /2\omega^2}
 e^{-\frac{m^2}{2\omega^2} \Big( x-\frac{1}{2} -\frac{m_{q_1}^2-m_{q_2}^2} {2m^2} \Big)^2 }\,,
\end{eqnarray}
with $N_m$ being a normalization factor such that
\begin{equation}
\int d^2p_{\rm T} \int_0^1 dx \mathbf{\Phi}_m^2 =1 \,,
\end{equation}
and $m_{q_1}$ and $m_{q_2}$ the constituent quark masses of the non-spectator and spectator quarks participating in the quark decaying process. We use $\omega=0.46\pm 0.05$~GeV and the following constituent quark masses in the model calculation: $m_u=m_d=0.33$~GeV, $m_s=0.50$~GeV, $m_b=4.9$~GeV. The value of $\omega$, which determines the average transverse quark momentum and is approximately the same for mesons with the same light spectator quark \cite{BSW}, is fixed by the  $\Br(\barB^0 \to \overline{K}_2^{0}\gamma)$ data. The numerical results for $\zeta_\perp^{K_J^{(*)}}(0) $ and $\zeta_\parallel^{K_J^{(*)}}(0)$ are collected in Table~\ref{knstmass}.

The detailed results for the branching fractions for $B \to \knpst \gamma$ decays are given in \tbl\ref{knstgamma}. Note that the decay with a heavier meson in the final state has a smaller branching fraction not only due to the smaller phase space and $\zeta_\perp^{K^{(*)}_J}(0)$ but also to the Clebsch-Gordan coefficient $\betatJ$ which is smaller for a larger spin $J$ (see \tbl\ref{alphabeta}). We find
\begin{eqnarray}
 && {\cal B}(B\to K^*(1410) \gamma) > {\cal B}(B\to K^*(1680)  \gamma)
 > {\cal B}(B\to K^*_2(1430) \gamma)
  \nonumber\\
 >&& {\cal B}(B\to K^*_2(1980) \gamma)
 > {\cal B}(B\to K^*_3(1780)\gamma) > {\cal B}(B\to K^*_4(2045) \gamma) \nonumber\\
 >&& {\cal B}(B\to K^*_5(2380) \gamma),
 \end{eqnarray}
and
 \begin{eqnarray}
 && {\cal B}(B\to K_1(1650) \gamma) > {\cal B}(B\to K_2(1820)  \gamma)
 \gtrsim {\cal B}(B\to K_2(1770) \gamma)  \nonumber\\
 >&& {\cal B}(B\to K_2(2250) \gamma)
 > {\cal B}(B\to K_3(2320)\gamma) > {\cal B}(B\to K_4(2500) \gamma) \nonumber\\
 >&& {\cal B}(B\to K_5(2600?) \gamma).
 \end{eqnarray}
It is interesting to note that we obtain $1.5 {\cal B}(B^- \to K^*(1680)\gamma) \sim {\cal B}(B^- \to K^*(1410)\gamma) =(27.2\pm 8.3) \cdot 10^{-6}$, whereas Ali, Ohl, and Mannel \cite{Ali:1992zd} found $7 {\cal B}(B^- \to K^*(1680)\gamma) \sim {\cal B}(B^- \to K^*(1410)\gamma) \simeq (35\pm 7) \cdot 10^{-6}$.

%%%%%%%%%%%%%%%%%%%%%%%%%%%%%%%%%%%%%%%%%%%%%%%%
\begin{table}[tbp]
\caption{The branching fractions of the $B \to \knpst\gamma$ decays in units of $10^{-6}$, where the errors are mainly due to the uncertainties of form factors. The corresponding photon energies in the $B$ rest frame are given in the last column.
}\label{knstgamma}
\begin{ruledtabular}
\begin{tabular}{lcc|rrr}
$\knpst$
& $J^{PC}$
& $n\,{}^{2S+1}L_J$
& $\Br(B^- \to \knpstm \gamma)$
& $\Br(\barBz \to \barknpstz \gamma) $
& $E_\gamma^B$ [\GeV]
\\
\hline
$\knstv{}{1410}$
& $1^{--}$
& $2\, {}^3S_1$?
& $27.2\pm 8.3$
& $25.0\pm 7.7$
& $2.45$
\\
$\knstv{}{1680}$
& $1^{--}$
& $1\,{}^3D_1$
& $17.8\pm 8.2$
& $16.4\pm 7.6$
& $2.36$
\\
$\knstv{2}{1430}$
& $2^{++}$
& $1\,{}^3P_2$
& $13.5\pm 4.1$
& $12.4\pm 3.8$
& $2.45$
\\
$\knstv{2}{1980}$
& $2^{+?}$
& $1\,{}^3F_2$ or $2\,{}^3P_2$?
& $5.5\pm 3.1$
& $5.1\pm 2.9$
& $2.27$
\\
$\knstv{3}{1780}$
& $3^{--}$
& $1\,{}^3D_3$
& $4.3\pm 2.1$
& $3.9\pm 1.9$
& $2.34$
\\
$\knstv{4}{2045}$
& $4^{++}$
& $1\,{}^3F_4$
& $1.4\pm 0.8$
& $1.3\pm 0.8$
& $2.24$
\\
$\knstv{5}{2380}$
& $5^{-?}$
& $1\,{}^3G_5$
& $0.4\pm 0.3$
& $0.3\pm 0.3$
& $2.10$
\\
%%%%%%%%%%%%%%%%%%%%%%%%%%%%%%%%%%%%%
$\knv{1}{1650}$
& $1^{+?}$
& $2\,{}^1P_1$ or $2\,{}^3P_1$?
& $18.3\pm 8.4$
& $16.9\pm 7.8$
& $2.38$
\\
%$\knv{2}{1580}$
%&
%&
%& $12.8^{+2.9}_{-2.6}$
%& $11.8^{+2.7}_{-2.4}$
%& $2.40$
%\\
$\knv{2}{1770}$
& $2^{-+}$
& $1\,{}^1D_2$
& $8.0\pm 3.9$
& $7.4\pm 3.6$
& $2.34$
\\
$\knv{2}{1820}$
& $2^{--}$
& $1\,{}^3D_2$?
& $8.5\pm 3.9$
& $7.9\pm 3.6$
& $2.33$
\\
$\knv{2}{2250}$
& $2^{-?}$
& $2\,{}^1D_2$
& $3.0\pm 2.2$
& $2.8\pm 2.0$
& $2.16$
\\
$\knv{3}{2320}$
& $3^{+?}$
& $1\,{}^1F_3$ or $1\,{}^3F_3$?
& $1.4\pm 1.1$
& $1.3\pm 1.0$
& $2.13$
\\
$\knv{4}{2500}$
& $4^{-?}$
& $1\,{}^1G_4$ or $1\,{}^3G_4$?
& $0.5\pm 0.4$
& $0.5\pm 0.3$
& $2.05$
\\
$\knv{5}{2600?}$
& $5^{+?}$
& $1\,{}^1H_5$ or $1\,{}^3 H_5$?
& $0.2\pm 0.2$
& $0.2\pm 0.2$
& $2.00$
\\ \hline
Total\footnote{We have assumed that
$\Br(B \to 2\,{}^1P_1 \gamma) \simeq \Br(B \to 2\,{}^3P_1\gamma)$
 if $2\,{}^1P_1$ and $2\,{}^3P_1$ states do not mix.
Analogously, we also assume that
$\Br(B \to 1\,{}^3F_2 \gamma) \approx \Br(B \to 2\,{}^3P_2 \gamma)$,
$\Br(B \to 1\,{}^1F_3 \gamma) \approx \Br(B \to 1\,{}^3F_3 \gamma)$,
$\Br(B \to 1\,{}^1G_4 \gamma) \approx \Br(B \to 1\,{}^3G_4 \gamma)$ and
$\Br(B \to 1\,{}^1H_4 \gamma) \approx \Br(B \to 1\,{}^3H_4 \gamma)$.
The summation of the branching fractions should be independent of the mixture due to the unitarity. Here we do not include decays involving $1\,{}^3G_3$ and $1\,{}^3 H_4$ states.}
&
&
& $135.9\pm 18.9$
& 125.2$\pm$ 17.4
\end{tabular}
\end{ruledtabular}
\end{table}
%%%%%%%%%%%%%%%%%%%%%%%%%%%%%%%%%%%%%%%%%%%%%%%

The total branching fractions of radiative $B$ meson decays involving resonant strange mesons\footnote{We do not include decays involving $1\,{}^3G_3$ and $1\,{}^3H_4$ states.} listed in Table~\ref{knstgamma}, together with
$\Br(B \to K^*(892)\gamma,K_1(1270)\gamma,K_1(1400)\gamma)$ \cite{Coan:1999kh,Nakao:2004th,:2008cy,Hatanaka:2008xj}, are
\begin{eqnarray}
\sum_{J=1}^5 \Br(\barBz \to \barknpstz \gamma; E_\gamma^B \gtrsim 2.0\,\GeV)
&=& (237^{+40}_{-34}) \times 10^{-6},
\\
\sum_{J=1}^5 \Br(B^- \to \knpstm \gamma; E_\gamma^B \gtrsim 2.0\,\GeV)
&=& (252^{+44}_{-36}) \times 10^{-6},
\end{eqnarray}
where $E_\gamma^B$ is the photon energy in the $B$ rest frame. Our result may hint at that the total branching fraction for the radiative $B$ decays with (nonresonant) two-body or three-body hadronic final states is about $100\times 10^{-6}$ (see also Ref.~\cite{Barberio:2008fa}), compared to the inclusive $B \to X_s\gamma$ data \cite{Chen:2001fja,Aubert:2005cua,Abe:2008sxa}
\begin{eqnarray}
\Br(B \to X_s \gamma; E_\gamma^B > 1.7\,\GeV) = (352\pm25) \times 10^{-6}.
\label{inclusive-gamma}
\end{eqnarray}

The $q^2$-dependence of form factors can be estimated by using the QCD counting rules \cite{Chernyak:1983ej,Hatanaka:2009gb}. We consider the Breit frame, where the $B$ meson and final state $\knpst$ move in the opposite directions but with the same magnitude of the momentum. In the large recoil region, where $q^2 \sim 0$, since the two quarks in mesons have to interact strongly with each other to turn around the spectator quark, the transition amplitude is dominated by the one-gluon exchange between the quark-antiquark pair and is therefore proportional to $1/E^2$. Thus we get $\langle
K_J^*(p_{K_J^*},\pm 1)|V^\mu|B(p_B)\rangle \propto \eps^{\mu\nu\rho\sigma}
p_{B\nu} p_{K_2^* \rho} \veps_{(J)}^*(\pm )_\sigma \times 1/E^2 $ and  $\langle
K_J^*(p_{K_J^*},0)|A^\mu|B(p_B)\rangle \propto p_{K_J^*}^\mu \times 1/E^2$. Consequently, we have $\zeta_{\perp,\parallel}(q^2)\sim 1/E^2$ in the large
recoil region. In other words, we can obtain that approximate forms:
$\zeta_{\perp,\para}^{\knpst}(q^2) =\zeta_{\perp,\para}^{\knpst}(0) \cdot (1-q^2/m_B^2)^{-2}$. This result is consistent with that obtained by Charles, Yaouanc, Oliver, P\`{e}ne and Raynal \cite{Charles:1998dr}. They used the light-cone sum rule method to show that the $B \to V$ LEET parameters satisfy $1/E^2$ scaling law, where $V\equiv$ vector meson. Essentially, their result is also suitable for the present case.

%%%%%%%%%%%%%%%%%%%%%%%%%%%%%%%%%%%%%%%%%%%%%%%%%%%%%%%%%%%%%%%%%%%%%%%%%%%%%%%%%%%%
\subsection{$B \to \knpst \ell^+ \ell^-$ Decays}
%%%%%%%%%%%%%%%%%%%%%%%%%%%%%%%%%%%%%%%%%%%%%%%%%%%%%%%%%%%%%%%%%%%%%%%%%%%%%%%%%%%%

The decay amplitude for $\barB \to \barknst \l^+\l^-$ is given by\footnote{For the amplitudes of $\barB \to \barkn \lpm$ decays, perform the following substitutions:
$V^{\knst} \to A^{\kn}$, $A_i^{\knst} \to V_i^{\kn}$ and $T_i^{\knst} \to T_i^{\kn}$. The result for the decay amplitude for $\barB \to \bar K^*(892) \l^+\l^-$ can be found in Ref.~\cite{Ali:1999mm}.
} \cite{Hatanaka:2009gb}
\begin{eqnarray}
\calM &=&
-i \frac{G_F \alphaem}{2\sqrt{2} \pi} V_{ts}^* V_{tb}^{} m_B
\left[ \calT^{\knst}_\mu \sbar \gamma^\mu b
     + \calU^{\knst}_\mu \sbar \gamma^\mu\gamma_5 b
\right],
\end{eqnarray}
where
\begin{eqnarray}
\calT^{\knst}_\mu &=&
 \calA^{(\knst)} \eps_{\mnrs} \tvepsst^{\nu} p_B^\rho p_{\knst}^\sigma
 -i m_B^2 \calB^{(\knst)} \tvepsst_\mu
 +i \calC^{(\knst)} (\tvepsst \cdot p_B) p_\mu
 +i \calD^{(\knst)} (\tvepsst \cdot p_B) q_\mu ,
 \nonumber\\&&
\\
\calU^{\knst}_\mu &=&
 \calE^{(\knst)} \eps_{\mnrs} \tvepsst^{\nu} p_B^\rho p_{\knst}^\sigma
 -i m_B^2 \calF^{(\knst)} \tvepsst_\mu
 +i \calG^{(\knst)} (\tvepsst \cdot p_B) p_\mu
 +i \calH^{(\knst)} (\tvepsst \cdot p_B) q_\mu,
 \nonumber\\&&
\end{eqnarray}
with $q_\mu \equiv p_B - p_{\knst}$.
The $\calD^{(\knst)}$-term vanishes when equations of motion of leptons are taken
into account.
The building blocks, $\calA^{(\knst)},\cdots$, and $\calH^{(\knst)}$ are given by
\begin{eqnarray}
\calA^{(\knst)} &=& \frac{2}{1 + \hatmknpst} c_9^{\eff}(\hats) V^{\knst}(s) + \frac{4\hatmb}{\hats} c_7^{\eff} T_1^{\knst}(s),
\\
\calB^{(\knst)} &=& (1+\hatmknpst) \left[
 c_9^{\eff}(\hats)  A_1^{\knst}(s) + 2 \frac{\hatmb}{\hats} (1-\hatmknpst)
 c_7^{\eff} T_2^{\knst}(s)
\right],
\\
\calC^{(\knst)} &=& \frac{1}{1-\hatmknpst} \biggl[
 (1-\hatmknpst) c_9^\eff(\hats) A_2^{\knst}(s) + 2\hatmb c_7^\eff
 \left(T_3^{\knst}(s) + \frac{1-\hatmknpst}{\hats}T_2^{\knst}(s)\right)
\biggr],\nonumber\\&&
\\
\calD^{(\knst)} &=& \frac{1}{\hats}
\biggl[
 c_9^\eff(\hats) \{(1+\hatmknpst) A_1^{\knst}(s) - (1-\hatmknpst)A_2^{\knst}(s)\}
\nonumber\\&&
\phantom{MMM} - 2\hatmknpst A_0^{\knst}(s)
 - 2\hatmb c_7^\eff T_3^{\knst}(s)
\biggr],
\\
\calE^{(\knst)} &=& \frac{2}{1+\hatmknpst} c_{10} V^{\knst}(s),
\quad
\calF^{(\knst)} = (1 + \hatmknpst) c_{10} A_1^{\knst}(s),
\\
\calG^{(\knst)} &=& \frac{1}{1+\hatmknpst} c_{10} A_2^{\knst}(s),
\\
\calH^{(\knst)} &=& \frac{1}{\hats} c_{10}\left[
 (1+\hatmknpst)A_1^{\knst}(s) - (1-\hatmknpst) A_2^{\knst}(s)
 - 2\hatmknpst A_0^{\knst}(s)
\right],
\end{eqnarray}
where $\hats = s/m_B^2$,
$\hatmknst=\mknst/m_B$,
$\hatmb = m_b/m_B$ and
$c_9^\eff(\hat s) = c_9 + Y_{\rm pert}(\hat s) + Y_{\rm LD} (\hat s)$ with the  perturbative $Y_{\rm pert}(\hat s)$ and long-distance $Y_{\rm LD}(\hat s)$  corrections \cite{Buras:1994dj,Ali:1991is,Lim:1988yu}.
$Y(\hats)_{\rm LD}$ involves $B \to \knst V(\cbar c)$ resonances, where $V(\cbar c)$ are the vector charmonium states \cite{Ali:1991is,Lim:1988yu}
\begin{eqnarray}
Y_{\rm LD}(\hats)
&=&
 - \frac{3\pi}{\alphaem^2} c_0
\sum_{V = \psi(1s),\cdots} \kappa_V \frac{\hat m_V \Br(V\to
\l^+\l^-)\hat{\Gamma}_{\rm tot}^V}{\hat s - \hat m_V^2 + i \hat m_V
\hat{\Gamma}_{\rm tot}^V},
\label{c9-LD}
\end{eqnarray}
with $\hat{\Gamma}_{\rm tot}^V \equiv \Gamma_{\rm tot}^V/\mB$. The relevant parameters can be found in Ref.~\cite{Hatanaka:2008gu}.

\begin{figure}[tbp]
\caption{Decay distributions of $\barBz \to\barknpstz \mupm$ decays.
The processes involving the confirmed $\knpst$ are plotted.
Solid [red],
dashed [orange],
dotted [green],
dot-dashed [blue], and
double-dot-dashed [black] curves from up to down correspond to
$\knpst=\knstv{}{1680},~\knstv{2}{1430},\knv{2}{1770},~\knstv{3}{1780}$, and $\knstv{4}{2045}$, respectively.
The thick and thin curves stand for the decay widths with and without charmonium resonances, respectively (see Eq.~\eqref{c9-LD}).
}\label{brplots}
\includegraphics{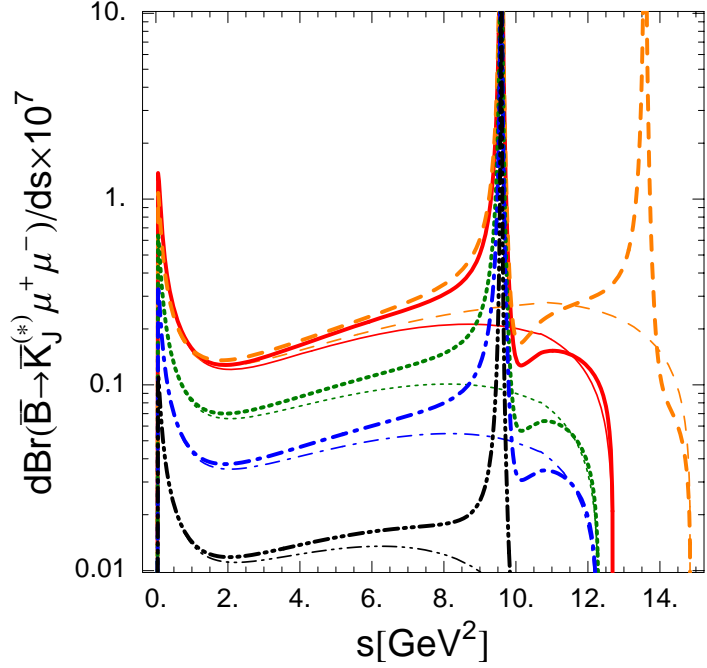}
\end{figure}

The longitudinal, transverse, and total differential decay widths are respectively given by
\begin{eqnarray}
\frac{d\Gamma_L}{d \hats}
\equiv\left. \frac{d\Gamma}{d \hats} \right|_{\substack{\alphal=\alphalJ\\ \betat=0}},
\quad
\frac{d\Gamma_T}{d \hats}
\equiv\left. \frac{d\Gamma}{d \hats} \right|_{\substack{\alphal=0\\ \betat=\betatJ}},
\quad
\frac{d\Gamma_{\rm total}}{d \hats}
\equiv\left. \frac{d\Gamma}{d \hats} \right|_{\substack{\alphal=\alphalJ\\ \betat=\betatJ}},
\end{eqnarray}
with
\begin{eqnarray}
\frac{d\Gamma}{d \hats}
&=&
\frac{G_F^2 \alphaem^2 m_B^5}{2^{10}\pi^5}|V_{ts}^*V_{tb}|^2\nonumber\\&&
\times \Biggl\{
%%%%%%%%%%%%%%%%%%%%%%%%%%%%%%%%%%%%%%%%%%%
\frac{1}{6}\left|\calA^{(\knst)}\right|^2 \hatus \hats \betat^2
\left\{
 3 \left[ 1- 2 (\hatmknst^2 + \hats) + (\hatmknst^2 - \hats)^2
 \right] - \hatus^2
\right\}
\nonumber\\&&
%%%%%%%%%%%%%%%%%%%%%%%%%%%%%%%%%%%%%%%%%%%
+\betat^2  \left|\calE^{(\knst)}\right|^2 \hats \frac{\hatus^3}{3}
\nonumber\\&&
%%%%%%%%%%%%%%%%%%%%%%%%%%%%%%%%%%%%%%%%%%%
+ \frac{1}{12 \hatmknst^2 \lambda} \left|\calB^{(\knst)}\right|^2 \hatus
\left\{ 3
 \left[ 1 - 2(\hatmknst^2 + \hats) + (\hatmknst^2 - s)^2 \right]
 - \hatus^2 \right\}
\nonumber\\&& \phantom{MMMM} \times
\left[ (-1 + \hatmknst^2 + \hats)^2 \alphal^2 + 8 \hatmknst^2 \hats \betat^2
\right]
\nonumber\\&&
%%%%%%%%%%%%%%%%%%%%%%%%%%%%%%%%%%%%%%%%%%%%
+ \frac{1}{12 \mknst^2 \lambda} \left|\calF^{(\knst)}\right|^2 \hatus
\Bigl\{
3\alphal^2 \lambda^2 \nonumber\\&& \phantom{MMMM}
+ \hatus^2
 \left[
16 \hatmknst^2 \hats \betat^2
- (1 - 2(\hatmknst^2+\hats) + \hatmknst^4
 + \hats^2 - 10\hatmknst^2 \hats) \alphal^2
\right]
\Bigr\}
\nonumber\\&&
%%%%%%%%%%%%%%%%%%%%%%%%%%%%%%%%%%%%%%%%%%%%
+ \alphal^2 \hatus \frac{\lambda}{4 \hatmknst^2}
\biggl[ \left|\calC^{(\knst)}\right|^2 \left(\lambda - \frac{\hatus^2}{3}\right)
+ \left|\calG^{(\knst)}\right|^2 \left(\lambda - \frac{\hatus^2}{3}
 + 4 \hatml^2 (2 + 2 \hatmknst^2 - \hats)\right)\biggr]
\nonumber\\&&
%%%%%%%%%%%%%%%%%%%%%%%%%%%%%%%%%%%%%%%%%%%%
- \alphal^2 \hatus \frac{1}{2 \hatmknst^2}   \biggl[
 \Re(\calB^{(\knst)} \calC^{(\knst)*}) \left(\lambda - \frac{\hatus^2}{3}\right)(1 - \hatmknst^2 - \hats)
\nonumber\\&&
\phantom{MMMM} +
 \Re(\calF \calG^*)\left\{
 \left(\lambda - \frac{\hatus^2}{3}\right)(1 - \hatmknst^2 - \hats)
 + 4 \hatml^2 \lambda \right\}
\biggr]
\nonumber\\&&
%%%%%%%%%%%%%%%%%%%%%%%%%%%%%%%%%%%%%%%%%%%
-2 \alphal^2 \hatus \frac{\hatml^2}{\hatmknst^2} \lambda \left[
   \Re(\calF^{(\knst)} \calH^{(\knst)*})
 - \Re(\calG^{(\knst)} \calH^{(\knst)*})(1 - \hatmknst^2)\right]
\nonumber\\&&
%%%%%%%%%%%%%%%%%%%%%%%%%%%%%%%%%%%%%%%%%%%
+ \alphal^2 \hatus \frac{\hatml^2}{\hatmknst^2} \hats \lambda  \left|\calH^{(\knst)}\right|^2
\Biggr\}.
\end{eqnarray}
Here $\hatu \equiv u/ m_B^2$ and $\hatus \equiv u(s) / m_B^2$,
where $u = - u(s) \cos\theta$,
\begin{eqnarray}
u(s) &\equiv& \sqrt{\lambda \left(1 - \frac{4\hatml^2}{\hats} \right)},
\\
\lambda &\equiv& 1 + \hatmknst^4 + \hats^2 - 2\hatmknst^2 - 2\hats - 2\hatmknst^2 \hats,
\end{eqnarray}
and $\theta$ is the angle between the moving directions of $\l^+$ and $B$ meson
in the center of mass frame of the $\l^+ \l^-$ pair. We show the decay distributions $d\Br(\barBz\to\barknpstz\mupm)/ds$ in \fig\ref{brplots} and summarize the corresponding branching fractions in \tbl\ref{knstll}. Because the decays involving heavier $K$-resonances have the smaller phase spaces and LEET form factors and because the Clebsch-Gordan coefficients, $\alpha_L^{(J)}$ and $\beta_T^{(J)}$, are smaller for a larger spin $J$, we obtain the following salient features:
 \begin{eqnarray}
 && {\cal B}(B\to K^*(1410) \mu^+\mu^-) > {\cal B}(B\to K^*_2(1430) \mu^+\mu^-)  > {\cal B}(B\to K^*(1680)  \mu^+\mu^-)   \nonumber\\
 >&& {\cal B}(B\to K^*_2(1980) \mu^+\mu^-)
 \approx {\cal B}(B\to K^*_3(1780)\mu^+\mu^-) > {\cal B}(B\to K^*_4(2045) \mu^+\mu^-) \nonumber\\
 >&& {\cal B}(B\to K^*_5(2380) \mu^+\mu^-),
 \end{eqnarray}
and
 \begin{eqnarray}
 && {\cal B}(B\to K_1(1650) \mu^+\mu^-) > {\cal B}(B\to K_2(1770) \mu^+\mu^-)  > {\cal B}(B\to K_2(1820)  \mu^+\mu^-)   \nonumber\\
 >&& {\cal B}(B\to K_2(2250) \mu^+\mu^-)
 > {\cal B}(B\to K_3(2320)\mu^+\mu^-) > {\cal B}(B\to K_4(2500) \mu^+\mu^-) \nonumber\\
 >&& {\cal B}(B\to K_5(2600?) \mu^+\mu^-).
 \end{eqnarray}

%%%%%%%%%%%%%%%%%%%%%%%%%%%%%%
\begin{figure}[tbhp]
\caption{Longitudinal fraction distributions $dF_L/ds$ of $\barB \to \barknpst \mupm$ decays as functions of $s$.
Solid [red],
dashed [orange],
dotted [green],
dot-dashed [blue] and
double-dot-dashed [black] curves stand for
$\knpst = \knstv{{}}{1680}$,
$\knstv{2}{1430}$,
$\knv{2}{1770}$,
$\knstv{3}{1780}$ and
$\knstv{4}{2045}$,
respectively.
}\label{fl-plot}
\includegraphics[width=3.5in]{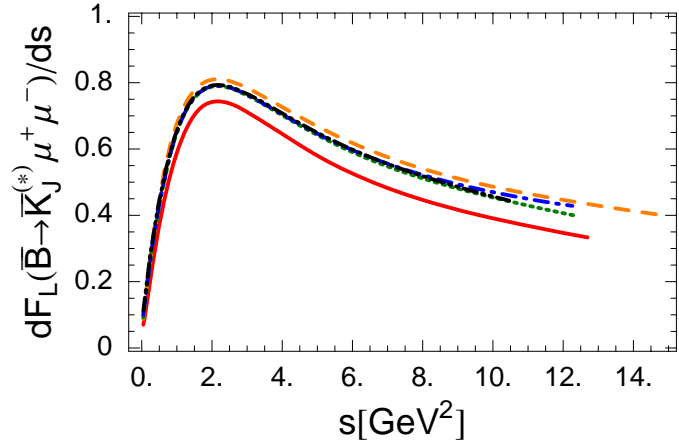}
\end{figure}

%%%%%%%%%%%%%%%%%%%%%%%%%%%%%%
%%%%%%%%%%%%%%%%%%%%%%%%%%%%%%
\newcommand{\sbl}[4]{{#1}\,{}^{#2}{#3}_{#4}}
\begin{table}[tbp]
\caption{Same as \tbl\ref{knstgamma} except for nonresonant  branching fractions of
$\barB \to \barknpst \mupm$ decays in units of $10^{-7}$.}\label{knstll}
\begin{ruledtabular}
\begin{tabular}{lllll}
& $J^{PC}$
& $n\,{}^{2S+1} L_J$
&$\Br(\barBz \to \barknpstz \mu^+ \mu^-) $
&$\Br(B^- \to \knpstm \mu^+ \mu^-)$
\\
\hline
$\knstv{}{1410}$
& $1^{--}$
& $\sbl{2}{3}{S}{1}$
& $5.4\err{+1.6}{-1.4}$
& $5.8\err{+1.7}{-1.5}$
\\
$\knstv{}{1680}$
& $1^{--}$
& $\sbl{1}{3}{D}{1}$
& $2.3 \err{+0.8}{-0.7}$
& $2.4 \err{+0.9}{-0.8}$
\\
$\knstv{2}{1430}$
& $2^{++}$
& $\sbl{1}{3}{P}{2}$
& $3.1\err{+0.9}{-0.8}$
& $3.3 \err{+1.0}{-0.9}$
\\
$\knstv{2}{1980}$
& $2^{+?}$
& $\sbl{1}{3}{F}{2}$ or $\sbl{2}{3}{P}{2}$
& $0.6\err{+0.3}{-0.2}$
& $0.6\err{+0.3}{-0.2}$
\\
$\knstv{3}{1780}$
& $3^{--}$
& $\sbl{1}{3}{D}{3}$
& $0.6 \err{+0.2}{-0.2}$
& $0.6 \err{+0.2}{-0.2}$
\\
$\knstv{4}{2045}$
& $4^{++}$
& $\sbl{1}{3}{F}{4}$
& $0.1\err{+0.1}{-0.1}$
& $0.2\err{+0.1}{-0.1}$
\\
$\knstv{5}{2380}$
& $5^{-?}$
& $\sbl{1}{3}{G}{5}$
& $0.03 \err{+0.02}{-0.01}$
& $0.03 \err{+0.02}{-0.01}$
%%%%%%%%%%%
\\
$\knv{1}{1650}$
& $1^{+?}$
& $\sbl{2}{1}{P}{1}$ or $\sbl{2}{3}{P}{1}$
& $2.6\err{+0.9}{-0.8}$
& $2.7\err{+1.0}{-0.8}$
\\
$\knv{2}{1770}$
& $2^{-+}$
& $\sbl{1}{1}{D}{2}$
& $1.1\err{+0.4}{-0.3}$
& $1.2\err{+0.4}{-0.4}$
\\
$\knv{2}{1820}$
& $2^{--}$
& $\sbl{1}{3}{D}{2}$?
& $0.9\err{+0.4}{-0.3}$
& $1.0\err{+0.4}{-0.3}$
\\
$\knv{2}{2250}$
& $2^{-?}$
& $\sbl{2}{1}{D}{2}$
& $0.2\err{+0.1}{-0.1}$
& $0.2\err{+0.1}{-0.1}$
\\
$\knv{3}{2320}$
& $3^{+?}$
& $\sbl{1}{1}{F}{3}$ or $\sbl{1}{3}{F}{3}$
& $0.1\err{+0.1}{-0.1}$
& $0.1\err{+0.1}{-0.1}$
\\
$\knv{4}{2500}$
& $4^{-?}$
& $\sbl{1}{1}{G}{4}$ or $\sbl{1}{3}{G}{4}$
& $0.03\err{+0.02}{-0.02}$
& $0.03\err{+0.02}{-0.02}$
\\
$\knv{5}{2600?}$
& $5^{+?}$
& $\sbl{1}{1}{H}{5}$ or $\sbl{1}{3}{H}{5}$?
& $0.01\err{+0.01}{-0.01}$
& $0.01\err{+0.01}{-0.01}$
\\
\hline
Total\footnote{Same as \tbl\ref{knstgamma}.}
&
&
& $19.9\err{+6.2}{-4.6}$
& $21.4\err{+7.1}{-5.2}$
\end{tabular}
\end{ruledtabular}
\end{table}
%%%%%%%%%%%%%%%%%%%%%%%%%%%%%%
%
%%%%%%%%%%%%%%%%%%%%%%%%%%%%%%%%%%%%%%%%%%%%%%%%%%%%%%%%%%%
% longitudinal fractions
%%%%%%%%%%%%%%%%%%%%%%%%%%%%%%%%%%%%%%%%%%%%%%%%%%%%%%%%%%%
In \fig\ref{fl-plot}, we plot the longitudinal fraction distributions for the $\barB \to \barknpst \mupm$ decays,
where
\begin{eqnarray}
\frac{d F_L}{d s} \equiv \frac{d\Gamma_L}{ds} \bigg/ \frac{d \Gamma_{\rm total}}{ds}.
\end{eqnarray}
Our result indicates that the longitudinal fraction distribution $dF_L/ds$ about $0.8$ at $s = 2 \, \GeV^2$, which also apply to the inclusive process. It is interesting to note that, for the new-physics models
with the flipped sign solution for $c_7^\eff$, $dF_L/ds$ can be reduced to be $\sim 0.6$ at $s = 2 \, \GeV^2$.

%%%%%%%%%%%%%%%%%%%%%%%%%%%%%
%%%%%%%%%%%%%%%%%%%%%%%%%%%%%

\begin{figure}[tbp]
\caption{Normalized forward-backward asymmetries for $\barB\to\barknpst\mupm$ decay.
Legends are the same as \fig\ref{fl-plot}.}
\label{afbplots}
\includegraphics{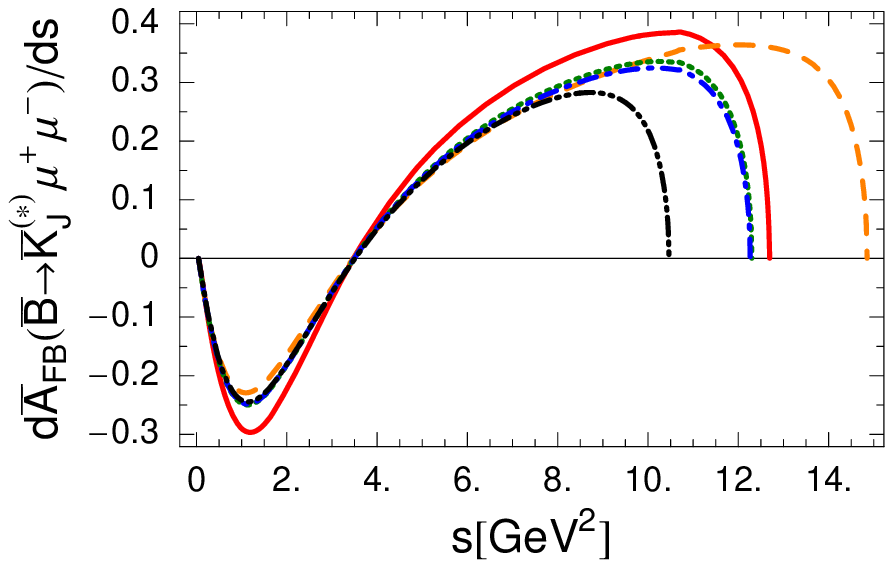}
\end{figure}

%%%%%%%%%%%%%%%%%%%%%%%%%%%%%%%%%%%%%%%%%%%%%%%%%%%%%%%%%%%
% forward-backward asymmetry
%%%%%%%%%%%%%%%%%%%%%%%%%%%%%%%%%%%%%%%%%%%%%%%%%%%%%%%%%%%
The forward-backward asymmetry of $\barB \to \barknst \l^+ \l^-$ is given by
\begin{eqnarray}
\frac{d A_{FB}}{d\hat{s}} &=&
 -  \left(\betatJ \right)^2 \frac{G_F^2\alphaem^2 m_B^5}{2^{10}\pi^5}
|V_{ts}^* V_{tb}|^2
\hat{s} \hat{u}(s)^2
\left[
   \Re \left( \calB^{(\knst)} \calE^{(\knst)*} \right)
 + \Re \left( \calA^{(\knst)} \calF^{(\knst)*} \right)
\right]
\nonumber \\
&=&
 -  \left( \betatJ \right)^2 \frac{G_F^2\alphaem^2 m_B^5}{2^{10}\pi^5}
|V_{ts}^* V_{tb}|^2
\hat{s} \hat{u}(s)^2
\nonumber\\&& \phantom{ii}
\times
\biggl[
\Re \left[ c_{10}^{} c_9^{\eff}(\hats) \right] V^{\knst} A_1^{\knst}
\nonumber\\&&\phantom{MM}
+ \frac{\hatm_b}{\hats} \Re( c_{10}^{} c_7^{\eff})
 \left\{
  (1-\hatmknst) V^{\knst} T_2^{\knst} + (1+ \hatmknst) A_1^{\knst} T_1^{\knst}
 \right\}
\biggr].
\end{eqnarray}
In \fig\ref{afbplots} we plot the normalized forward-backward asymmetry $d\bar{A}_{FB}/ds \equiv (dA_{FB}/ds)/(d\Gamma_{\rm total}/ds)$.
\newcommand{\gtrkn}[2]{\gtrapprox s_0^{\knv{#1}{#2}}}
\newcommand{\gtrknst}[2]{\gtrapprox s_0^{\knstv{#1}{#2}}}
Using the form factors in Eqs.~\eqref{LEETrelationA0}-\eqref{LEETrelationT3}, we can easily obtain the forward-backward asymmetry zero, $s_0$, satisfying
\begin{eqnarray}
\Re \left[ c_9^{\eff}(\hats_0) c_{10} \right] &=& -2 \frac{\hatm_b}{\hats_0}
\Re(c_7^{\eff} c_{10}) \frac{1-\hats_0}{1+\hatm_{\knpst}^2 - \hats_0}.
\label{zerocond}
\end{eqnarray}
We note that $s_0$ is independent of the form factors but depends only on $\mknpst$. Under the variation of $\hatm_{\knpst}^2$, we get
\begin{eqnarray}
\delta \hats_0 \simeq
 \frac{(\hats_0 - 1)\hats_0}{(\hats_0-1)^2 + \hatm_{\knpst}^2} \delta \hatm_{\knpst}^2,
\end{eqnarray}
or
\begin{eqnarray}
\delta s_0 &\simeq& - s_0 \cdot \frac{\delta m_{\knpst}^2}{m_B^2}.
\label{s0-variation}
\end{eqnarray}
Since $\delta m_{\knpst}^2 \ll s_0$ and $m_{\knpst}^2 \ll m_B^2$,
we thus expect the following relation in the SM:
\begin{eqnarray}
s_0^{K^*(980)} \approx 3.5\, \GeV^2
\gtrknst{}{1410}
\gtrknst{2}{1430}
\gtrknst{}{1680}
\gtrkn{2}{1770}
\gtrknst{3}{1780}
\gtrkn{2}{1820}
\nonumber\\
\gtrknst{2}{1980}
\gtrknst{4}{2045}
\gtrkn{2}{2250}
\gtrkn{3}{2320}
\gtrknst{5}{2380}
\gtrkn{4}{2500}
\gtrkn{5}{2600?}
.
\nonumber\\
\end{eqnarray}

%%%%%%%%%%%%%%%%%%%%%%%%%%%%%%%%%%%%%%
%%%%%%%%%%%%%%%%%%%%%%%%%%%%%%%%%%%%%%

%%%%%%%%%%%%%%%%%%%%%%%%%%%%%%%%%%%%%%%%%%%%%%%%%%%%%%%%%%%%%%%%%%%%%%
\subsection{$\barB \to \barknpst \nu \nubar$ Decays}
\begin{figure}[tbp]
\caption{The $d\Br(\barB\to\barknpst \nu\nubar)/d q^2$ as functions of the missing invariant mass-squared $q^2$.
The solid [black], dashed [blue], dotted [green] and dot-dashed [red] curves correspond to the total decay rate
and the polarization rates with helicities $h=0,-1,+1$, respectively.
}\label{nunuplot}
\includegraphics[width=0.45\linewidth]{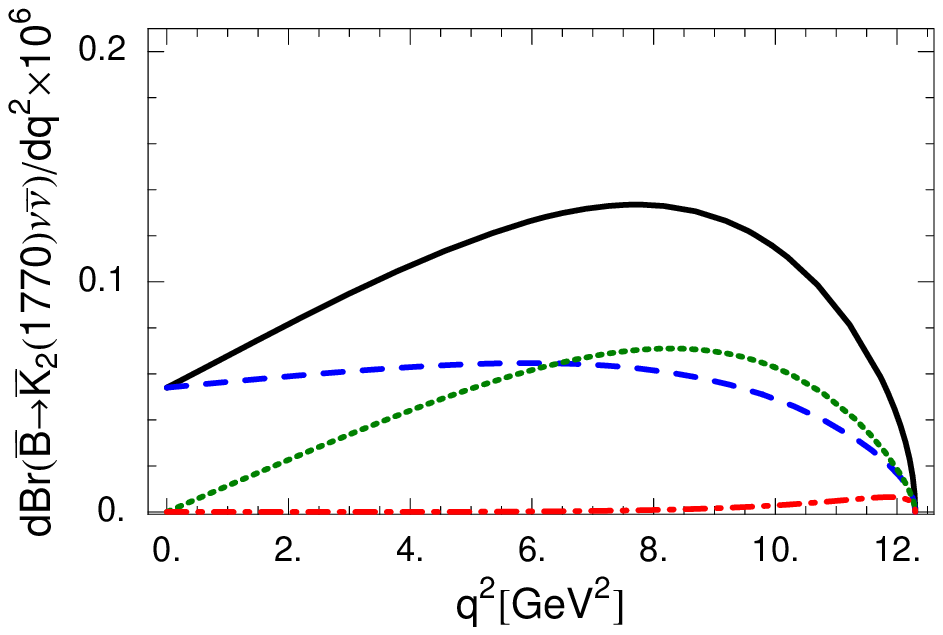}
\includegraphics[width=0.45\linewidth]{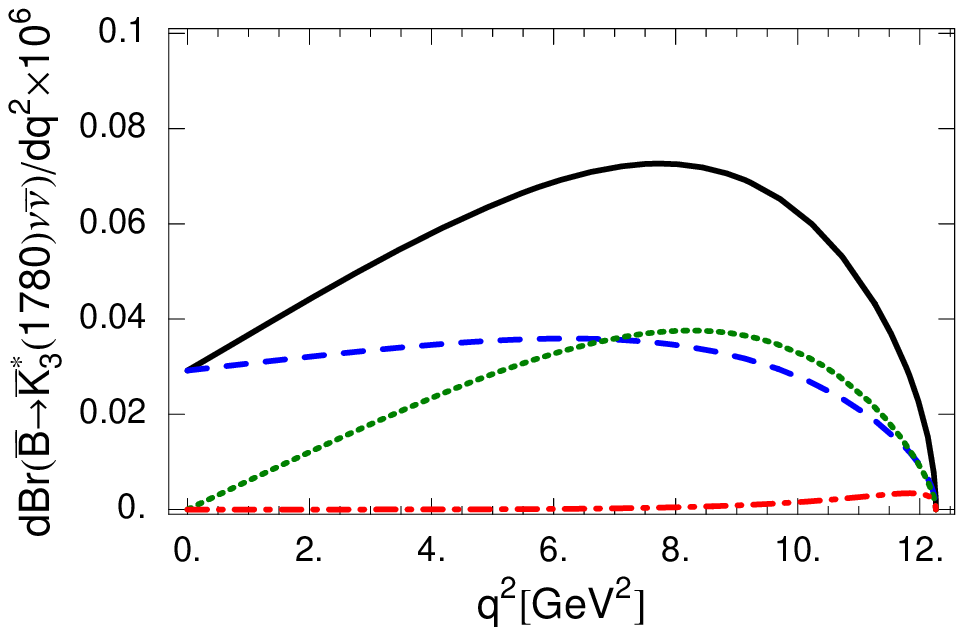}\\
\includegraphics[width=0.45\linewidth]{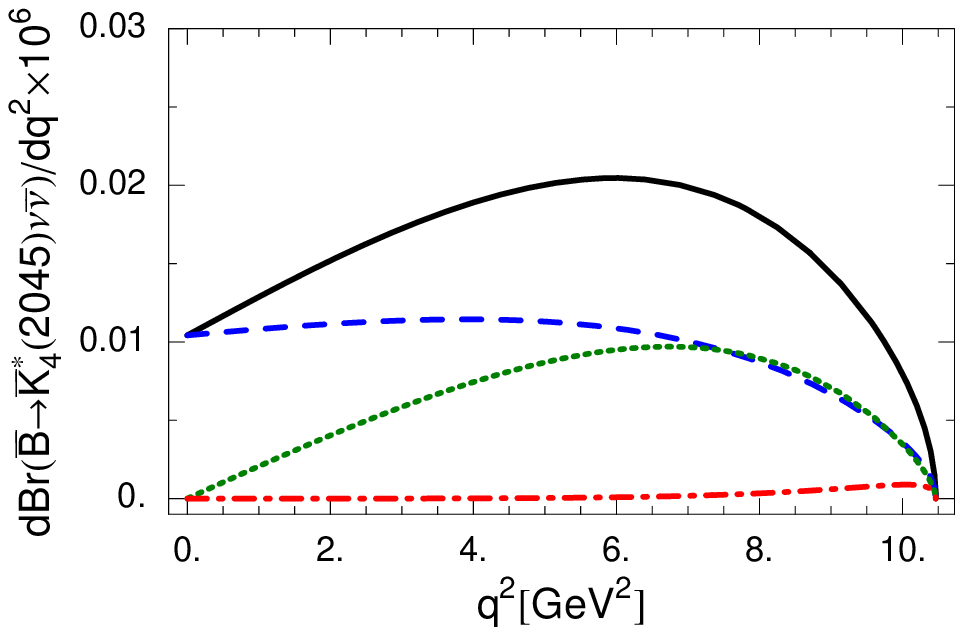}
\includegraphics[width=0.45\linewidth]{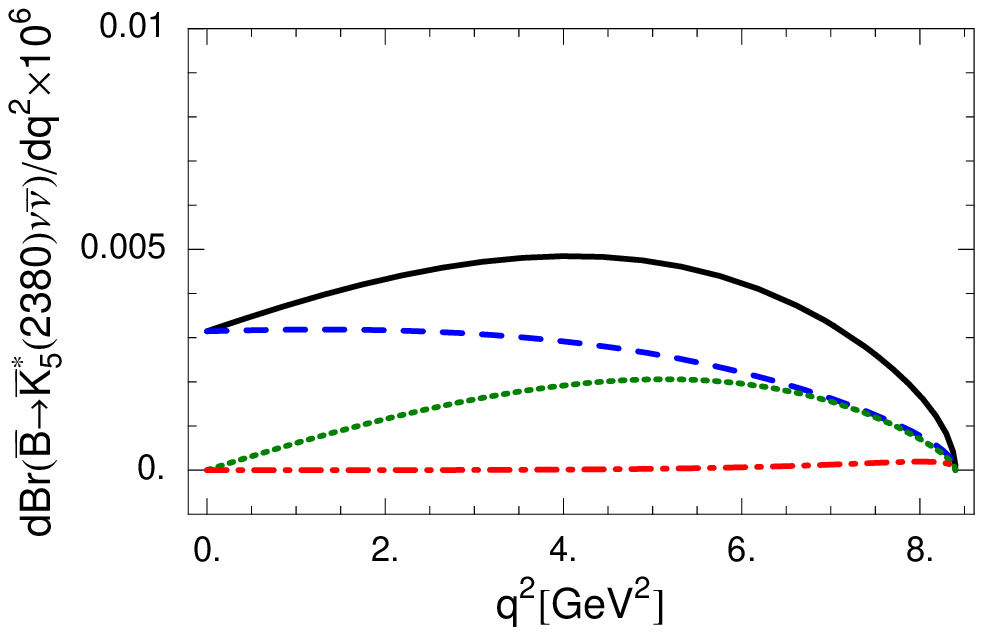}
\end{figure}

%%%%%%%%%%%%%%%%%%%%%%%%%%%%%%%%%%%%%%%%%%%%%%%%%%%%%%%%%%%%%%%%%%%%%%
The effective weak Hamiltonian relevant to the $\barB \to \barknpst\nu\nubar$ decays
are given by
\begin{eqnarray}
\calH_{\eff} =
c_L \sbar \gamma^\mu(1-\gamma_5) b\, \nubar \gamma_\mu (1-\gamma_5) \nu +
c_R \sbar \gamma^\mu(1+\gamma_5) b\, \nubar \gamma_\mu (1-\gamma_5) \nu +
\Hc,
\end{eqnarray}
where $c_L$ and $c_R$ are coefficients for left- and right-handed weak hadronic currents, respectively.
In the SM, $c_R^{SM} = 0$ and
\begin{eqnarray}
c_L^{SM} &=&
\frac{G_F}{\sqrt{2}} \frac{\alpha_{EM}}{2\pi\sin^2\theta_W}
V_{tb}^{} V_{ts}^* X(x_t) = 2.9 \times 10^{-9},
\end{eqnarray}
where the detailed form of $X(x_t)$ has been given in Refs.~\cite{Inami:1980fz,Buchalla:1993bv}.
The missing invariant mass-squared distributions, corresponding to polarizations $h=0,\pm1$ of the final $\barknst$ for $\barB\to\barknst\nu\nubar$ decays are\footnote{For the amplitudes of $\barB \to \barkn \nu\nubar$ decays, perform the following replacements: $V^{\knst} \to A^{\kn}$, $A_i^{\knst} \to V_i^{\kn}$.},
\begin{eqnarray}
\frac{d\Gamma_0}{d q^2} &=& 3 \left( \alphalJ \right)^2
\frac{|\vec{p}'|}{48\pi^3}  \frac{|c_L-c_R|^2}{\mknst^2}
 \nonumber\\&&
\times \left[
 (m_B + \mknst)(m_B E' - \mknst^2) A_1^{\knst}(q^2)
- \frac{2m_B^2}{m_B + \mknst}|\vec{p}'|^2 A_2^{\knst}(q^2)
\right]^2,
%\nonumber\\
\\
%%%
\frac{d\Gamma_{\pm1}}{d q^2}
&=& 3 \left( \betatJ \right)^2
\frac{|\vec{p}'| q^2}{48\pi^3}
\nonumber\\&&
 \times \Biggl|
 (c_L + c_R) \frac{2 m_B |\vec{p}'|}{m_B + \mknst} V^{\knst}(q^2)
\mp
 (c_L - c_R) (m_B + \mknst) A_1^{\knst}(q^2)
 \Biggr|^2
,
%\nonumber\\
\end{eqnarray}
where  the factor $3$ counts the numbers of the neutrino generations,
$(E',\vec{p}')$ is the $\barknst$ energy-momentum in the $B$-meson rest frame,
and $q^2$ is the invariant mass squared of the neutrino-antineutrino pair with
$0 \le q^2 \le (m_B - \mknst)^2$.
In \fig\ref{nunuplot}, we show the differential distributions as functions of the missing invariant mass squared in the SM.
The results for branching fractions are summarized in \tbl\ref{knstnunu}. At $q^2=0$, where the neutrino and antineutrino are nearly collinear in the $B$ rest frame, the decay is predominated by the zero helicity amplitude. Moreover, as expected from the left-handed $b_L \to s_L$ transition in the SM, $d\Gamma_+/dq^2$ is always suppressed at least by $(m_s/m_b)^2$, compared with $d\Gamma_0/dq^2$ and $d\Gamma_-/dq^2$. We obtain the relation: $d\Gamma_0/dq^2>d\Gamma_-/dq^2 \gg  d\Gamma_+/dq^2$.

%
%%%%%%%%%%%%%%%%%%%%%%%%%%%%%%
%%%%%%%%%%%%%%%%%%%%%%%%%%%%%%
%
%%%%%%%%%%%%%%%%%%%%%%%%%%%%%%
\begin{table}[tbp]
\caption{The branching fractions of the
$B \to \knpst \nu\nubar$ decays in units of $10^{-6}$.
The first and second errors correspond to the uncertainties of the form factors
$\zeta_\perp^{\knpst}$ and $\xi^{\knpst}$, respectively.
}\label{knstnunu}
\begin{ruledtabular}
\begin{tabular}{lllll}
& $J^{PC}$
& $\sbl{n}{2S+1}{L}{J}$
&$\Br(\barBz \to \barknpstz \nu \nubar)$
&$\Br(B^- \to \knpstm \nu \nubar)$
\\
\hline
$\knstv{}{1410}$
& $1^{--}$
& $\sbl{2}{3}{S}{1}$?
& $4.3\err{+1.3}{-1.1}$
& $4.6\err{+1.4}{-1.2}$
\\
$\knstv{}{1680}$
& $1^{--}$
& $\sbl{1}{3}{D}{1}$
& $1.8\err{+0.7}{-0.6}$
& $2.0\err{+0.7}{-0.6}$
\\
$\knstv{2}{1430}$
& $2^{++}$
& $\sbl{1}{3}{P}{2}$
& $2.5\err{+0.7}{-0.6}$
& $2.6\err{+0.8}{-0.7}$
\\
$\knstv{2}{1980}$
& $2^{+?}$
& $\sbl{1}{3}{F}{2}$ or $\sbl{2}{3}{P}{2}$?
& $0.4\err{+0.2}{-0.2}$
& $0.5\err{+0.2}{-0.2}$
\\
$\knstv{3}{1780}$
& $3^{--}$
& $\sbl{1}{3}{D}{3}$
& $0.5\err{+0.2}{-0.2}$
& $0.5\err{+0.2}{-0.2}$
\\
$\knstv{4}{2045}$
& $4^{++}$
& $\sbl{1}{3}{F}{4}$
& $0.11\err{+0.05}{-0.04}$
& $0.11\err{+0.06}{-0.05}$
\\
$\knstv{5}{2380}$
& $5^{-?}$
& $\sbl{1}{3}{G}{5}$
& $0.02\err{+0.01}{-0.01}$
& $0.02\err{+0.01}{-0.01}$
%%%%%%%%%%%%%%%%
\\
$\knv{1}{1650}$
& $1^{+?}$
& $\sbl{2}{1}{P}{1}$ or $\sbl{2}{3}{P}{1}$?
& $2.1\err{+0.7}{-0.6}$
& $2.2\err{+0.8}{-0.7}$
\\
$\knv{2}{1770}$
& $2^{-+}$
& $\sbl{1}{1}{D}{2}$
& $0.9\err{+0.3}{-0.3}$
& $0.9\err{+0.4}{-0.3}$
\\
$\knv{2}{1820}$
& $2^{--}$
& $\sbl{1}{3}{D}{2}$?
& $0.7\err{+0.3}{-0.2}$
& $0.8\err{+0.3}{-0.3}$
\\
$\knv{2}{2250}$
& $2^{-?}$
& $\sbl{2}{1}{D}{2}$
& $0.2\err{+0.1}{-0.1}$
& $0.2\err{+0.1}{-0.1}$
\\
$\knv{3}{2320}$
& $3^{+?}$
& $\sbl{1}{1}{F}{3}$ or $\sbl{1}{3}{F}{3}$
& $0.07\err{+0.05}{-0.03}$
& $0.07\err{+0.05}{-0.04}$
\\
$\knv{4}{2500}$
& $4^{-?}$
& $\sbl{1}{1}{G}{4}$ or $\sbl{1}{3}{G}{4}$
& $0.02\err{+0.02}{-0.01}$
& $0.02\err{+0.01}{-0.01}$
\\
$\knv{5}{2600}$
& $5^{+?}$
& $\sbl{1}{1}{H}{5}$ or $\sbl{1}{3}{H}{5}$
& $0.008\err{+0.006}{-0.005}$
& $0.008\err{+0.007}{-0.005}$
\\
\hline
Total\footnote{Same as \tbl\ref{knstgamma}.}
&
&
& $16.2\err{+4.1}{-3.0}$
& $17.3\err{+4.7}{-3.5}$
%%%%%%%%%%
\end{tabular}
\end{ruledtabular}
\end{table}
%%%%%%%%%%%%%%%%%%%%%%%%%%%%%

%%%%%%%%%%%%%%%%%%%%%%%%%%%%%%%%%%%%%%%%%%%%%%%%%%%%%%%%%%%%%%%%%%%%%%%%
\section{Summary}\label{sec:summary}

We have formulated $B \to \knpst$ form factors using large energy effective theory techniques. We have studied the radiative and semileptonic $B$ decays involving the higher strange resonance $\knpst$ in the final state. The main results are as follows.

\begin{itemize}
\item
The transition form factors in the large recoil region can be represented in terms of two independent LEET form factors, $\zeta^{\knpst}_\perp(q^2)$ and $\zeta^{\knpst}_\para(q^2)$. According to the QCD counting rules, these two form factors exhibit the dipole $q^2$ dependence in the large recoil region (and in the LEET limit). We have further estimated $\zeta^{\knpst}_\perp(0)$ and $\zeta^{\knpst}_\para(0)$ in the BSW model.

\item
The branching fractions for decays $\barB\to\barknpst\gamma$, $\barB\to\barknpst\lpm$ and $\barB\to\barknpst\nu\nubar$ with higher $K$-resonances are suppressed due to the smaller phase space and $\zeta^{\knpst}_{\perp,\parallel}$, and/or due to the smaller Clebsch-Gordan coefficients, $\betatJ$ and $\alphalJ$, in case of larger spin-$J$.

\item
We find that for $\barB \to \barknpst \lpm$ decays,
the longitudinal fraction distribution $dF_L/ds \simeq 0.8$ at $s = 2 \, \GeV^2$, and the forward-backward asymmetry zero $s_0 \approx 3.5\, \GeV^2$.
The asymmetry zero is independent of the form factors in the LEET limit and highly insensitive to $\mknpst$.

\item
For the $\barB \to \barknpst \nu \bar\nu$ decay,  the branching fraction is predominated by the zero helicity amplitude at $q^2=0$, where the neutrino and antineutrino are nearly collinear in the $B$ rest frame. As expected from the left-handed $b_L \to s_L$ current in the SM, $d\Gamma_+/dq^2$ is always suppressed at least by $(m_s/m_b)^2$, compared with $d\Gamma_0/dq^2$ and $d\Gamma_-/dq^2$. We thus predict the relation: $d\Gamma_0/dq^2>d\Gamma_-/dq^2 \gg  d\Gamma_+/dq^2$.

\end{itemize}

%%%%%%%%%%%%%%%%%%%%%%%%%%%%%%%%%%%%%%%%%%%%%%%%%%%%%%%%%%%%%%%%%%%%%%%%
\begin{acknowledgments}
This research was supported in parts by the National Science Council of R.O.C. under Grant No.~NSC96-2112-M-033-004-MY3 and No.~NSC97-2811-033-003 and by the National Center for Theoretical Science.

\end{acknowledgments}

\appendix

%%%%%%%%%%%%%%%%%%%%%%%%%%%%%%%%%%%%%%%%%%%%%%%%%%%%%%%%%%%%%%%%%%%%%%%%%%%%%%%%%%%%%%%
\section{$\barB \to \barkn$ form factors}\label{app:btotn-FF}
%%%%%%%%%%%%%%%%%%%%%%%%%%%%%%%%%%%%%%%%%%%%%%%%%%%%%%%%%%%%%%%%%%%%%%%%%%%%%%%%%%%%%%%
$\barB \to \barkn$ transition form factors in the LEET limit are given by
\begin{eqnarray}
\bra{\barkn}|A^\mu|\ket{\barB}
&=& -i 2 E \left(\frac{\mkn}{E}\right)^{J-1} \zetaaA_\perp \eps^{\mnrs}
 v_\nu n_\rho e^*_\sigma,
\label{LEETFF-vectorA}
\\
%%%%%%%%%%%%%%%%%%%%%%%%%%%%
\bra{\barkn}|V^\mu|\ket{\barB}
&=& 2 E \left(\frac{\mkn}{E}\right)^{J-1} \zetavA_\perp
\left[ e^{*\mu} - (e^* \cdot v) n^\mu \right]
\nonumber\\&&
+ 2 E \left(\frac{\mkn}{E}\right)^{J} (e^* \cdot v)
 \left[\zetavA_{\para} n^\mu + \zetavA_{\para,1} v^\mu \right],
\label{LEETFF-axialvectorA}
\\
%%%%%%%%%%%%%%%%%%%%%%%%%%%%
\bra{\barkn}|T_A^{\mu\nu}|\ket{\barB}
&=& -2 E \left(\frac{\mkn}{E}\right)^{J}  \zetatfA_\para (e^* \cdot v)
 \eps^{\mnrs} v_\rho n_\sigma
\nonumber\\&&
- 2 E \left(\frac{\mkn}{E}\right)^{J-1}\zetatfA_\perp \eps^{\mnrs} n_\rho [e^*_\sigma
 - (e^* \cdot v) n_\sigma]
\nonumber\\&&
- 2 E \left(\frac{\mkn}{E}\right)^{J-1}\zetatfA_{\perp,1} \eps^{\mnrs} v_\rho [e^*_\sigma
 - (e^* \cdot v) n_\sigma],
\label{LEETFF-tensorA}
\\
%%%%%%%%%%%%%%%%%%%%%%%%%%%%
\bra{\barkn}|T^{\mu\nu}|\ket{\barB}
&=& i 2 E \left(\frac{\mkn}{E}\right)^{J-1} \zetatA_{\perp,1}
 \left\{
 \left[ e^{*\mu} - (e^* \cdot v)n^\mu\right] v^\nu - (\mu \leftrightarrow \nu)
 \right\}
\nonumber\\&&
+i 2 E \left(\frac{\mkn}{E}\right)^{J-1} \zetatA_{\perp} \left\{
 [e^{*\mu} - (e^* \cdot v)n^\mu] n^\nu - (\mu \leftrightarrow \nu)
\right\}
\nonumber\\&&
+i 2 E \left(\frac{\mkn}{E}\right)^{J} \zetatA_{\para}
(e^* \cdot v) (n^\mu v^\nu - n^\nu v^\mu),
\label{LEETFF-axialtensorA}
\end{eqnarray}
where $\mkn$ is the mass of the $\kn$.
$\bra{\barkn}|T_A^{\mu\nu}|\ket{\barB}$ is related to $\bra{\barkn}|T^{\mu\nu}|\ket{\barB}$ by the relation: $\sigma^{\mu\nu}\eps_{\mnrs}= 2i \sigma^{\rho\sigma} \gamma_5$.
From operator relations Eqs.~\eqref{relation-scalar}-\eqref{relation-axialtensor} and
\begin{eqnarray}
\bar{s}_n \gamma_5 b_v &=& - n_\mu \bar{s}_n \gamma^\mu \gamma_5 b_v,
\end{eqnarray}
we obtain
\begin{eqnarray}
\zetavA_\perp = \zetaaA_\perp = \zetatA_\perp = \zetatfA_\perp &\equiv& \zeta^{\kn}_\perp,
\\
\zetaaA_\para = \zetatA_\para = \zetatfA_\para &\equiv& \zeta^{\kn}_\para,
\\
\zetaaA_{\para,1} = \zetatfA_{\perp,1} = \zetatA_{\perp,1} &=& 0,
\end{eqnarray}
and thus find that there are only two independent form factors, $\zeta^{\kn}_\perp(q^2)$ and $\zeta^{\kn}_{\para}(q^2)$.

$\barB \to \barkn$ form factors are given by
\begin{eqnarray}
\bra{\barkn(p_{\kn},\lambda)}|\sbar \gamma^\mu \gamma_5 b|\ket{\barB(p_B)}
&=&
 - i \frac{2}{m_B + \mkn} \tilde{A}^{\kn}(q^2)
\eps^{\mnrs} p_{B\nu} p_{\kn \rho} e(\lambda)_\sigma^*,
\\
%%%%%%%%%%%%%%%%%%%%%%%%%%%%%%%%
\bra{\barkn(p_{\kn},\lambda)}|\sbar \gamma^\mu b|\ket{\barB(p_B)}
&=&
 2 \mkn \tilde{V}_0^{\kn}(q^2) \frac{e(\lambda)^* \cdot p_B}{q^2} q^\mu
\nonumber\\&&
+ \left(m_B + \mkn \right) \tilde{V}_1^{\kn}(q^2)
\left[ e(\lambda)^{*\mu} - \frac{e(\lambda)^* \cdot p_B}{q^2} q^\mu \right]
\nonumber\\&&
- \tilde{V}_2^{\kn}(q^2) \frac{e(\lambda)^* \cdot p_B}{m_B + \mkn}
\left[ p_B^\mu + p_{\kn}^\mu - \frac{m_B^2 - \mkn^2}{q^2} q^\mu
\right],
%\nonumber\\
\\
%%%%%%%%%%%%%%%%%%%%%%%%%%%%%%%%
\bra{\barkn(p_{\kn},\lambda)}|\sbar \sigma^{\mu\nu} \gamma_5 q_\nu b|\ket{\barB(p_B)}
&=&
2 \tilde{T}_{1}^{\kn} (q^2) \eps^{\mnrs} p_{B\nu} p_{\kn \rho} e(\lambda)^*_\sigma,
\\
%%%%%%%%%%%%%%%%%%%%%%%%%%%%%%%%
\bra{\barkn(p_{\knst},\lambda)}|\sbar \sigma^{\mu\nu}q_\nu b|\ket{\barB(p_B)}
 &=&
 i \tilde{T}_{2}^{\kn} (q^2) \left[
\left(m_B^2 - \mkn^2\right) e(\lambda)^{*\mu}
 - \left(e(\lambda)^* \cdot p_B\right) \left(p_B^\mu + p_{\kn}^\mu\right)
\right]
\nonumber\\&&
+i \tilde{T}_{3}^{\kn}(q^2) \left(e(\lambda)^* \cdot p_B\right)
\nonumber\\&& \phantom{MMM} \times
\left[ q^\mu - \frac{q^2}{m_B^2 - \mkn^2}\left(p_B^\mu + p_{\kn}^\mu\right)
\right].
%\nonumber\\
\end{eqnarray}
We can further obtain the following relations,
\begin{eqnarray}
\tilde{V}_0^{\kn}(q^2) \left(\frac{|\vec{p}_{\kn}|}{\mkn}\right)^{J-1}
\equiv
V_0^{\kn}(q^2) &\simeq& \left(1-\frac{\mkn^2}{m_B E}\right) \zeta_{\para}^{\kn}(q^2)
+ \frac{\mkn}{m_B} \zeta_{\perp}^{\kn}(q^2),
\\
\tilde{V}_1^{\kn}(q^2) \left(\frac{|\vec{p}_{\kn}|}{\mkn}\right)^{J-1}
\equiv
V_1^{\kn}(q^2) &\simeq& \frac{2 E}{m_B + \mkn} \zeta_{\perp}^{\kn}(q^2),
\\
\tilde{V}_2^{\kn}(q^2) \left(\frac{|\vec{p}_{\kn}|}{\mkn}\right)^{J-1}
\equiv
V_2^{\kn}(q^2) &\simeq& \left(1+\frac{\mkn}{m_B}\right) \left[
\zeta_{\perp}^{\kn}(q^2) -  \frac{\mkn}{E} \zeta_\para^{\kn}(q^2)
\right],
\\
\tilde{A}^{\kn}(q^2) \left(\frac{|\vec{p}_{\kn}|}{\mkn}\right)^{J-1}
\equiv
A^{\kn}(q^2)
&\simeq& \left(1+\frac{\mkn}{m_B}\right) \zeta_{\perp}^{\kn}(q^2),
\\
\tilde{T}_1^{\kn}(q^2) \left(\frac{|\vec{p}_{\kn}|}{\mkn}\right)^{J-1}
\equiv
T_1^{\kn}(q^2) &\simeq& \zeta_{\perp}^{\kn}(q^2),
\\
\tilde{T}_2^{\kn}(q^2) \left(\frac{|\vec{p}_{\kn}|}{\mkn}\right)^{J-1}
\equiv
T_2^{\kn}(q^2) &\simeq& \left(1-\frac{q^2}{m_B^2 - \mkn^2}\right)\zeta_{\perp}^{\kn}(q^2),
\\
\tilde{T}_3^{\kn}(q^2) \left(\frac{|\vec{p}_{\kn}|}{\mkn}\right)^{J-1}
\equiv
T_3^{\kn}(q^2) &\simeq& \zeta_{\perp}^{\kn}(q^2)
                - \left(1-\frac{\mkn^2}{m_B^2}\right)\frac{\mkn}{E} \zeta_\para^{\kn}(q^2),
\end{eqnarray}
where use of $p_3/E \simeq 1$ has been made.
Recalling that
\begin{eqnarray}
\tveps(0)^\mu = \alphalJ \veps(0)^\mu,
\quad
\tveps(\pm1)^\mu = \betatJ \veps(\pm1)^\mu,
\end{eqnarray}
we can easily generalize the studies for $B \to \knst \gamma$,
$B \to \knst \lpm$ and $B \to \knst \nu \nubar$ to
$B \to \kn\gamma$, $B \to \kn \lpm$ and $B \to \kn \nu\nubar$
by the following replacements:
\begin{eqnarray}
V^{\knst} \to A^{\kn},
\quad
A_i^{\knst} \to V_i^{\kn}\quad (i=0,1,2),
\quad
T_j^{\knst} \to T_j^{\kn}\quad (j=1,2,3).
\end{eqnarray}

%%%%%%%%%%%%%%%%%%%%%%%%%%%%%%%%%%%%%%%%%%%%%%%%%%%%%%%%%%%%%%%%%%%%%%%

\end{document}